\documentclass[11pt]{article}
\usepackage{comment,soul}
\usepackage[hidelinks]{hyperref}
\usepackage{enumitem}
\usepackage{amsmath,amssymb,epsf,cite,graphicx,subfigure}
\usepackage{braket,tikzsymbols}
\usepackage[bbgreekl]{mathbbol}
\usepackage{comment}
\usepackage[export]{adjustbox}
\usepackage{dsfont}
\usepackage{faktor}
\usepackage{mathrsfs}
\usepackage{float}
\usepackage{empheq}
\usepackage{tikz}

\usepackage{comment}

\numberwithin{equation}{section}

\definecolor{airforceblue}{rgb}{0.36, 0.54, 0.66}
\newcommand{\beq}{\begin{equation}}
\newcommand{\eeq}{\end{equation}}

\newcommand{\dd}{\text{d}}

\DeclareMathOperator{\Vol}{vol}

\begin{document}
\baselineskip=15.5pt
\pagestyle{plain}
\setcounter{page}{1}

\begin{center}
{\LARGE \bf Inner products and closed universes}
\vskip 1cm

\textbf{Jordan Cotler$^{1,a}$ and Kristan Jensen$^{2,b}$}

\vspace{0.5cm}

{\it ${}^1$ Department of Physics, Harvard University, Cambridge, MA 02138, USA \\}
{\it ${}^2$ Department of Physics and Astronomy, University of Victoria, Victoria, BC V8W 3P6, Canada\\}

\vspace{0.3cm}

{\tt  ${}^a$jcotler@fas.harvard.edu, ${}^b$kristanj@uvic.ca\\}

\medskip

\end{center}

\vskip1cm

\begin{center}
{\bf Abstract}
\end{center}
\hspace{.3cm} 

We study overlaps of late-time states in semiclassical de Sitter quantum gravity. In a family of deformations of de Sitter JT gravity, we compute the no-boundary wavefunction to one-loop and derive the ultralocal late-time measure. The one-universe contribution to the norm is exactly four times the sphere amplitude, so the pairing used by the Lorentzian measure differs from that implicit in Euclidean gravity. After summing over disconnected final universes, the no-boundary norm exponentiates the one-universe result, giving $\langle \!\langle \text{HH}|\text{HH}\rangle\!\rangle \approx \exp(4Z_{\rm sphere})$ rather than the sum over closed geometries, $Z_{\rm closed}\approx \exp(Z_{\rm sphere})$. We also review an analogous mismatch in Einstein gravity with positive cosmological constant, where the late-time no-boundary norm vanishes at one-loop, differing from the Euclidean sphere amplitude. We discuss the implications for cutting and gluing in perturbative gravity, and features that arise when extending this construction to sums over topologies.

\newpage

\tableofcontents

\section{Introduction}

In this paper we consider perturbative quantum gravity with positive cosmological constant in spacetimes with compact spatial slices. In this setting there is no ordinary Hilbert space of scattering states, and yet in a semiclassical regime at late time we can describe states by data on slices of large spatial volume. The basic question we study is: what is the overlap of these states? 

This question is relevant for theoretical cosmology, where its answer is an ingredient in the calculation of cosmological correlations~\cite{Maldacena:2002vr,Weinberg:2005vy,Arkani-Hamed:2018kmz,Baumann:2022jpr}, and in recent discussions of baby universes~\cite{Marolf:2020xie,Usatyuk:2024mzs,Harlow:2026hky}, observers~\cite{Chandrasekaran:2022cip,Kudler-Flam:2024psh,Chen:2024rpx,Speranza:2025joj,Abdalla:2025gzn,Harlow:2025pvj}, and ensemble interpretations of gravitational path integrals~\cite{Saad:2019lba,Pollack:2020gfa,Cotler:2020ugk,Marolf:2020xie}. Here we study a narrower, concrete problem, namely the semiclassical overlap of late-time states implied by perturbative gravity, or equivalently the completeness relation obeyed by these semiclassical states.

In ordinary quantum field theory, completeness relations are closely related to the cutting and gluing relations obeyed by path integrals. There, cutting and gluing is implemented quantum mechanically by introducing a resolution of the identity on a cut. One obtains transition amplitudes by fixing fields on the cut, and then composition follows by integrating over them. In the context of very late times in asymptotically de Sitter space, we can proceed by analogy and obtain a semiclassical completeness relation by integrating over asymptotically de Sitter boundary conditions modulo boundary gauge transformations~\cite{Chakraborty:2023yed,Chakraborty:2023los,Cotler:2025gui}. The resulting object is simply the ultralocal measure on the data of the conformal boundary of de Sitter space.

This semiclassical measure is the analogue of the path integral completeness relation in quantum field theory. We would like to understand what this measure computes, and especially how the norm of the no-boundary state obtained from this measure is related to the gravitational path integral over closed geometries. After all, the no-boundary state is prepared by summing over geometries that smoothly fill in late-time boundary data, and gluing this state to its conjugate appears pictorially to produce closed geometries.  Indeed, in perturbative quantum field theory on a fixed de Sitter background this intuition is correct. If we compute the norm of the Bunch--Davies state using the ultralocal measure on late-time field configurations, and compute the sphere partition function using the same regulator, normalization of the path-integral measure, and renormalization scheme, the two agree.  Although the result can be scheme dependent, the scheme dependence is identical in the norm and sphere computations (see e.g.~the appendix of~\cite{Cotler:2025gui} for a recent discussion).

The main result of this manuscript is that in quantum gravity, the no-boundary norm and the sum over closed geometries are different even in the loop expansion of semiclassical quantum gravity. That is, the natural semiclassical pairing does not reproduce the cutting-and-gluing relation implicit in the Euclidean gravitational path integral.

To do so we study a wide family of two-dimensional models, namely deformations of de Sitter Jackiw-Teitelboim (JT) gravity~\cite{Maldacena:2019cbz, Cotler:2019nbi, Dey:2025osp,Blommaert:2024whf}, described by an action
\beq
	S = \frac{S_0}{4\pi}\int \dd^2x \sqrt{-g} \,R + \frac{1}{16\pi G_2} \int \dd^2x \sqrt{-g}\left( \Phi R - V(\Phi)\right) + (\text{bdy})\,, \quad V = 2\Phi - \frac{U}{2}\Phi^2 + O(\Phi^3)\,.
\eeq
In this de Sitter context (unlike in AdS~\cite{Maxfield:2020ale, Witten:2020wvy}) we do not yet have control over the genus expansion with a nonlinear potential, so we work in a semiclassical regime with $G_2\ll 1$ and $U$ finite and positive. We evaluate the sphere amplitude $Z_{\rm sphere}$ to one-loop order, and from it the leading approximation to the sum over closed geometries, which is its exponential $\exp(Z_{\rm sphere})$.

We then compute the norm of the no-boundary state in this model using the semiclassical inner product. To do this, we first consider one-universe contributions to the no-boundary wavefunction and the measure, both to one-loop order. This one universe is a circle parameterized by $\theta \sim \theta + 2\pi$. The loop-corrected one-universe wavefunction is calculable at the leading order in the genus expansion, building on previous work in dS JT together with accounting for the quadratic potential $U$. The ultralocal measure is on a boundary einbein $e(\theta)$ and dilaton $\phi(\theta)$ modulo a diff$\times$Weyl redundancy. The no-boundary wavefunction has support when $\phi(\theta)$ has no roots, and then this measure collapses to $\dd\varphi/|\varphi|$ for a single real gauge-invariant variable $\varphi$, essentially the renormalized dilaton. The one-loop and one-universe contribution to the norm evaluated with this measure equals precisely four times the sphere amplitude.

Next we remove the projection onto the one-universe sector. The $n$-universe wavefunction of the no-boundary state is a tensor product at leading order in the genus expansion, $n$ copies of the single-universe wavefunction, and the $n$-universe measure is just $n$ copies of the one-universe measure up to a division by an $S_n$ permutation symmetry that exchanges the universes. This division contributes a factor of $1/n!$ to the $n$-universe norm, and for this reason the one-universe answer exponentiates. This inverse factor has a statistical interpretation whereby multiple universes act like identical bosons: the semiclassical overlap of $n$-universe states involves $n!$ different ways of connecting the $n$ circles in the ket to those in the bra. The semiclassical norm is then $\exp(4Z_{\rm sphere})$, which notably differs from the sum over disconnected spheres $\exp(Z_{\rm sphere})$.

Both of these one-loop computations depend on the careful division by diffeomorphisms in a theory of gravity. We can work in a common normalization of the path integral measure that locates the mismatch. In this normalization the nonexceptional bosonic and ghost modes cancel. The surviving fluctuation determinants come entirely from the $\ell=0,1$ sectors on the sphere and the Schwarzian sector of the no-boundary wavefunction. For the sphere we divide by the carefully normalized volume of the $SO(3)$ isometry and gauge-fix the $\ell=1$ conformal modes. The Lorentzian pairing instead contains a finite Jacobian from using a residual $SO(2,1)$ (rather than $SO(3)$) gauge symmetry to fix certain joint dilaton-metric zero modes of the late-time data. These finite-dimensional gauge factors differ, and their ratio produces the factor of four.

It is interesting to compare this result in dilaton gravity with the analogous one in pure Einstein gravity with positive cosmological constant in $d\geq 3$ dimensions, or in Einstein gravity coupled to stable matter. There, the semiclassically most likely outcome for a single universe in the no-boundary state is that the late-time universe is a round sphere. However, in the loop expansion, and using the same methods we use to compute the no-boundary norm in dilaton gravity, the round-sphere sector gives a vanishing contribution to the no-boundary norm. The culprit is an unfixed, residual $SO(d,1)$ gauge symmetry of the ultralocal measure on the late-time data, whereas the Euclidean sphere is finite and involves a division by the $SO(d+1)$ isometries. Unlike in dilaton gravity, boundary data cannot fix this symmetry. Although the bulk spacetime appearing in this overlap has the topology of a $d$-sphere, its contribution to the no-boundary norm differs from the Euclidean sphere amplitude, which is finite and moreover has Polchinski's phase~\cite{Polchinski:1988ua}, $Z_{\rm sphere} \propto i^{d+2}$. 

Our interpretation is that the Euclidean path integral implements a different pairing of the no-boundary wavefunction than that of the Lorentzian norm. Both constructions involve sphere topology and admit a cut on which one integrates over boundary data. This strongly suggests a common description in terms of complexified data on the cut. However the pairing relevant for the Euclidean sphere amplitude involves, at one-loop, a division by the sphere isometries, a division to which the Lorentzian pairing is blind. In forthcoming work with Shi and Turiaci~\cite{CTJSWIP1}, we show this is indeed the case at one-loop in deformed JT and Einstein gravity by pulling the late-time overlap back to a pairing on the equator of the Euclidean sphere, finding disagreement with the Euclidean path integral.

It is natural to ask whether the construction of an inner product extends beyond the semiclassical regime, for example through the genus expansion of dilaton gravity. The semiclassical answer suggests a candidate prescription. At leading order, the overlap of multi-universe states comes from a path integral over ``short'' cylinders that connect universes in the bra to those in the ket. This suggests extending the overlap by summing over connected ``short'' geometries that have at least one bra boundary and at least one ket boundary. Such a prescription would exclude components that cap off independently, or that connect only bra boundaries or only ket boundaries. Although the connectivity condition has a clear topological meaning, we do not know how to define ``short'' intrinsically once higher-genus or topology-changing geometries contribute.

The obstructions we find beyond the semiclassical regime are natural extensions of the one-loop mismatch: a pairing intrinsic to a cut does not know all of the redundancies of the closed geometry obtained after gluing. That is, a pairing intrinsic to a chosen cut naturally divides only by transformations that preserve that presentation, whereas the closed gravitational path integral divides by the full redundancies of the glued geometry, including large diffeomorphisms that move the cut. Moreover, the same closed manifold can arise from many decompositions into bra and ket components. In two-dimensional models we show that the sewing required to recover the closed amplitude depends on topology away from the cut, and that the usual symmetry factors do not remove the multiplicity of decompositions. These observations strongly suggest that the pairing implemented by the Euclidean path integral is state-dependent.

The candidate pairing suggested by the semiclassical overlap differs from Marolf–Maxfield-like constructions~\cite{Marolf:2020xie}, in which overlaps are defined by summing over all bulk fillings of specified boundary conditions. The difference is visible already for two-universe states. The semiclassical overlap contains the two ways of pairing the two bra circles with the two ket circles by short cylinders. An unrestricted sum over fillings also includes configurations in which boundaries cap off separately, or in which the two bra circles connect to each other while the two ket circles connect to each other. These are sensible gravitational amplitudes, but they are absent from the semiclassical completeness relation. More fundamentally, in Lorentzian de Sitter gravity the topology of a filling does not determine the quantum-mechanical object it computes. In dS JT gravity, a long cylinder connecting the far past to the far future is a transition amplitude defined with an $i\varepsilon$ prescription and does not equal the ultralocal short cylinder overlap. Pure dS$_3$ gravity gives a sharper version of the same feature, since the long global cylinder and the short overlap carry different effective anomaly coefficients. Thus an unrestricted sum over fillings does not by itself specify an inner product; a de Sitter version of such a construction must also retain the Lorentzian distinction between overlaps and transition amplitudes.

Taken together, these results largely answer the question posed at the beginning. In perturbative gravity, the ultralocal measure on asymptotic boundary data implements the semiclassical completeness relation. In dilaton gravity and in Einstein gravity, this pairing is simply different from that implemented by the Euclidean path integral. This does not mean that the ultralocal measure is wrong: it is the natural semiclassical completeness relation. What fails is the assumption that this completeness relation is the one used when cutting open the Euclidean gravitational path integral.

The rest of this manuscript is organized as follows. In Section~\ref{sec:test} we compute the sphere amplitude and leading no-boundary norm in dilaton gravity using the semiclassical overlap. We also review how the analogous computation works in Einstein gravity. We then comment on some aspects of this overlap beyond the semiclassical regime in Section~\ref{sec:obstructions}, and wrap up with a Discussion in Section~\ref{sec:discussion}.

\section{The semiclassical closed universe inner product}
\label{sec:test}

\subsection{Cutting and gluing on a slice}
\label{subsec:local-cutting-gluing}

In ordinary quantum field theory, cutting and gluing are implemented by a resolution of the identity on the cut.  If a closed spacetime $M$ is decomposed into two pieces $M_-$ and $M_+$ along a codimension-one slice $\Sigma$, then fixing the field configuration $\chi$ on $\Sigma$ gives the local sewing relation
\beq
Z_M = \int [\dd\chi]_\Sigma\, \Psi_{M_-}[\chi]\Psi_{M_+}[\chi]\,.
\eeq
When the two pieces are related by a reflection or orientation reversal so that one piece can be identified as a ket and the other a bra, this becomes
\beq
Z_M = \langle \Psi|\Psi\rangle = \int [\dd\chi]_\Sigma\, \Psi^*[\chi]\Psi[\chi]\,.
\eeq
Equivalently, in the path integral the introduction of a cut on $\Sigma$ is an insertion of the identity
\beq
\mathds{1}_{\Sigma} = \int [\dd\chi]_\Sigma |\chi\rangle\langle\chi|\,.
\eeq

In gravity the analogous statement is more delicate, even before summing over topologies.  Once a cut $\Sigma$ has been chosen, the data on the cut include the induced metric, together with any matter fields.  We denote these data schematically by $X = (\gamma,\chi)$.  Since these labels are gauge-redundant, the local gravitational analogue of the identity takes the form
\beq
\label{E:localGravityIdentity}
\mathds{1}_{\Sigma,\,{\rm loc}} = \int \frac{[\dd\gamma][\dd\chi]}{\text{gauge}}\, |\gamma,\chi\rangle\!\rangle \langle\!\langle\gamma,\chi|\,,
\eeq
where the measure includes the gauge-fixed Faddeev-Popov determinant and residual symmetry factors.  This expression should be regarded as a local semiclassical gluing rule for a specified cut $\Sigma$.  We emphasize that it is not, by itself, a non-perturbative prescription.

We would like to understand to what extent this cutting-and-gluing rule extends to gravity. The simplest arena where we can study this question involves the sphere. Semiclassically, a sphere can be cut into two disks. In what follows $|\text{HH}\rangle\!\rangle$ denotes the no-boundary state, including its decomposition into sectors with any number of connected late-time universes and, when included, its expansion in topology.  Let $\Pi_n$ denote the projector onto states where the late-time universe is a disjoint union of $n$ topological spheres $\mathbb{S}^{d-1}$. Thus $\Pi_1|\text{HH}\rangle\!\rangle$ is the one-sphere component of the no-boundary state. Operationally we interpret $\Pi_1$ as being given by the expression~\eqref{E:localGravityIdentity} where the cut $\Sigma$ is a single sphere, and similarly for $\Pi_n$. We would therefore like to understand if
\beq
\label{E:sphereNormTest}
	Z_{\rm sphere} \stackrel{?}{=} \langle\!\langle {\rm HH}|\Pi_1|{\rm HH}\rangle\!\rangle\,,
\eeq
is true. More generally, after including disconnected final universes, one may ask whether the no-boundary norm matches the sum over closed geometries.

The relation~\eqref{E:sphereNormTest}, which holds in quantum field theory, is non-trivial in perturbative gravity. Importantly, both the left- and right-hand sides of~\eqref{E:sphereNormTest} are independently defined. The local semiclassical inner product is defined independently by the ultralocal quotient over boundary data in~\eqref{E:localGravityIdentity}, or equivalently by the corresponding short-time gravitational path integral when such a derivation is available as in dS JT gravity~\cite{Cotler:2019nbi, Cotler:2019dcj, Cotler:2023eza, Cotler:2024xzz}. Equation~\eqref{E:sphereNormTest} instead tests whether the sphere amplitude admits a quantum mechanical interpretation in terms of the no-boundary state.  The distinction is important because a closed Euclidean path integral need not, in general, have the Hermiticity, positivity, or residual gauge quotient appropriate to a Lorentzian overlap.

In practice, we evaluate the overlap on the right-hand side of~\eqref{E:sphereNormTest} at late Lorentzian time rather than directly on an equatorial cut.  The disk path integral prepares a no-boundary wavefunction, we evolve to a regime where the asymptotic states and the ultralocal measure~\eqref{E:localGravityIdentity} are under control, and we then use the independently defined late-time inner product to glue the two wavefunctions. The question is whether this prescription reconstructs the closed sphere path integral, including the one-loop measure normalization.

In this Section we show that the answer is no in a controlled two-dimensional setting: deformations of de Sitter JT gravity, to one-loop order in the gravitational interaction. We compute the Euclidean sphere amplitude directly and compare it with the one-universe projection of the late-time no-boundary norm.  With the same ultralocal normalization of the path integral measure on the two sides, the two results have the same dependence on coupling constants to leading order in the genus expansion and at one loop, but differ by an overall factor of four. 

In~\cite{Cotler:2025gui} we previously computed the right-hand side of~\eqref{E:sphereNormTest} in Einstein gravity with positive cosmological constant, with and without perturbative scalar matter. There~\eqref{E:sphereNormTest} fails to hold: the late-time norm perturbatively vanishes on account of a noncompact conformal stabilizer of the late-time round sphere. In contrast, the Euclidean sphere amplitude is nonzero and has Polchinski's phase. We conclude that the Lorentzian pairing differs from that implicit in the Euclidean path integral already at one-loop both in dilaton and in Einstein gravity.

\subsection{The test in deformed de Sitter JT gravity}
\label{subsec:test-deformed-dsjt}

We now specialize to a class of two-dimensional models in which the test of the closed universe inner product can be performed explicitly.  As mentioned in the Introduction, we study deformations of dilaton gravity with Lorentzian action
\beq
	S =S_{\rm top} +  \frac{1}{16\pi G} \int \dd^2x \sqrt{-g} \left( \Phi R - V(\Phi)\right) + (\text{bdy})\,, \qquad V(\Phi) =2\Phi - \mathcal{U}(\Phi)\,,
\eeq
where $\mathcal{U}$ is a sum of monomials $\frac{1}{2}\mathcal{U}''(0)\Phi^2 + \frac{1}{3!}\mathcal{U}'''(0) \Phi^3 + \cdots$.  At $\mathcal{U} = 0$ this is exactly the dS version of JT gravity~\cite{Jensen:2016pah,Maldacena:2016upp,Engelsoy:2016xyb}. Here we take $\mathcal{U}''(0) = U\neq 0$ fixed, $G\ll 1$, and work to one-loop order in $G$ around the saddle $\Phi = 0$. In computing the late-time norm there is a scaling regime in which the late-time asymptotics are approximately those of dS JT gravity, allowing us to borrow from previous results for the classification of asymptotic states and the inner product thereof~\cite{Maldacena:2019cbz, Cotler:2019nbi, Cotler:2019dcj,Cotler:2023eza, Cotler:2024xzz}.

There are two reasons this family of theories is useful for the gluing test.  First, the Euclidean sphere calculation is completely controlled: the round sphere with constant dilaton at $\Phi = 0$ is a saddle, and the one-loop determinant can be computed directly.  Second, the same deformation leads to a nontrivial late-time probability distribution for the dilaton. In pure dS JT gravity the leading no-boundary wavefunction is a phase times a power of the late-time dilaton mode, but for $U > 0$ the deformation produces a right-sign Gaussian distribution. Thus the one-universe norm is finite and computable.

Both sides of the comparison depend on the normalization of the path integral measure. Rescaling the ultralocal path integral measure by a factor of $\alpha$ has the effect of additively renormalizing the topological coupling $S_0$ by $-\frac{3}{2}\ln \alpha$, and hence rescales both the sphere amplitude and the product of two disk wavefunctions by $\alpha^{-3}$. We therefore use the same ultralocal normalization of the bulk path integral measure in the sphere determinant and in the disk wavefunction.  With this common convention, the equality between the sphere amplitude and the one-universe norm is meaningful.

The calculation proceeds in two steps.  We first compute the Euclidean sphere amplitude directly.  We then compute the late-time one-universe norm of the Hartle-Hawking wavefunction and compare the two.  A closely related sphere computation appears in the minimal-string context~\cite{Anninos:2021ene,Mahajan:2021nsd}; here we perform the analogous calculation directly in the dilaton-gravity variables appropriate to the dS JT deformation.

\subsubsection{Sphere amplitude}
\label{subsubsec:deformedJT-sphere}

On a topological sphere the Euclidean action is
\beq
	S_E = -2S_0 +\frac{1}{16\pi G} \int \dd^2x \sqrt{g}(-\Phi R+V(\Phi))\,.
\eeq
The equations of motion are
\begin{align}
\begin{split}
	R - V'& = 0\,,
	\\
	(D_{\mu}D_{\nu} - g_{\mu\nu}D^2)\Phi - \frac{1}{2}g_{\mu\nu}V & = 0\,,
\end{split}
\end{align}
and the trace of the latter implies
\beq
	D^2\Phi = -V\,.
\eeq
There are round sphere solutions with constant dilaton at roots of $V$ with $V' > 0$, and there is a single such root at $\Phi = 0$ with $V'(0)=2$ so that the sphere has unit radius. The on-shell action is simply
\beq
	S_E = -2S_0\,.
\eeq

Now consider the spectrum of perturbations. Denote the metric on the round sphere as $g_{\mu\nu}$, and the covariant derivative formed from it as $\nabla_{\mu}$. Letting the metric be $g_{\mu\nu}+h_{\mu\nu}$, with $h = g^{\mu\nu}h_{\mu\nu}$, we fix the gauge
\beq
	\nabla^{\nu}h_{\mu\nu} - \frac{1}{2}\nabla_{\mu}h = 0\,.
\eeq
This choice does not completely fix the gauge symmetry: there is a residual gauge symmetry generated by conformal Killing vectors that we have to treat separately. We account for these residual modes later. Including the Lagrange multiplier field $\lambda^{\mu}$ and the Faddeev-Popov ghosts, the quadratic action with $\mathcal{U}''(0) = U$ reads
\begin{align}
\begin{split}
	S_E & = -2S_0 + \frac{1}{16\pi G}\int \dd^2x \sqrt{g}\left( \Phi(\nabla^2 +2)\omega - \frac{1}{2}U\Phi^2\right) +S_{\lambda} +  S_{\rm FP} + (\text{cubic})\,,
	\\
	S_{\lambda} &= \frac{i}{16\pi G}\int \dd^2x  \sqrt{g}\,\lambda^{\mu}\left(\nabla^{\nu}h_{\mu\nu} - \frac{1}{2}\nabla_{\mu}h\right) \,,
	\\
	S_{\rm FP} &= \frac{1}{16\pi G}\int \dd^2x \sqrt{g}\, b^{\mu}(\nabla^2 +1)c_{\mu}  \,.
\end{split}
\end{align}
Decomposing the most general 2d metric fluctuation as 
\beq
	h_{\mu\nu} = \omega g_{\mu\nu} + \nabla_{\mu}V_{\nu} + \nabla_{\nu}V_{\mu}\,, 
\eeq
we further decompose $(\Phi,\omega)$ into scalar harmonics\footnote{We pick a real orthonormal basis for the $Y_{\ell m}$'s, $\int d^2\Omega \,Y_{\ell m}Y_{\ell'm'} = \delta_{\ell \ell'}\delta_{mm'}$, so that the coefficients $\Phi_{\ell m}$, $\omega_{\ell m}$, and so on, are real.} 
\beq
	\Phi = \sum_{\ell=0}^{\infty}\sum_{m=-\ell}^{\ell}\Phi_{\ell m}Y_{\ell m}\,, \qquad \omega = \sum_{\ell=0}^{\infty}\sum_{m=-\ell}^{\ell}\omega_{\ell m}Y_{\ell m}\,,
\eeq 
the vector field $\lambda^{\mu}$ as
\begin{align}
\begin{split}
	\lambda^{\mu} &= \nabla^{\mu}\lambda + \varepsilon^{\mu\nu}\nabla_{\nu}\tilde{\lambda}\,,
	\\
	\lambda &= \sum_{\ell=1}^{\infty} \sum_{m=-\ell}^{\ell} \lambda_{\ell m}\frac{Y_{\ell m}}{\sqrt{\ell(\ell+1)}}\,, \qquad \tilde{\lambda} = \sum_{\ell= 1}^{\infty} \sum_{m=-\ell}^{\ell} \tilde{\lambda}_{\ell m}\frac{Y_{\ell m}}{\sqrt{\ell(\ell+1)}}\,, 
\end{split}
\end{align}
and similarly for the other vectors $(V_{\mu},b^{\nu},c_{\rho})$. This is equivalent to decomposing vectors into normalized vector harmonics. Rescaling the fluctuations by $\sqrt{16\pi G}$, the quadratic action becomes
\begin{align}
\begin{split}
	S_2 &=- \sum_{\ell=0}^{\infty}\sum_{m=-\ell}^{\ell}\left(  (\ell-1)(\ell+2)\Phi_{\ell m} \omega_{\ell m} + \frac{1}{2}U|\Phi_{\ell m}|^2\right)
	\\
	& \qquad - \sum_{\ell=2}^{\infty} \sum_{m=-\ell}^{\ell} (\ell-1)(\ell+2) \left( i \lambda_{\ell m} V_{\ell m} + i \tilde{\lambda}_{\ell m} \tilde{V}_{\ell m} - b_{\ell m} c_{\ell m} - \tilde{b}_{\ell m} \tilde{c}_{\ell m}\right)\,.
\end{split}
\end{align}
We normalize the ultralocal pairings on the rescaled fluctuations to be
\begin{align}
\begin{split}
\label{E:ultralocal}
	(\Phi_1,\Phi_2) & = \frac{1}{2\pi}\int \dd^2x \sqrt{g}\,\Phi_1\Phi_2\,, \qquad (h_1,h_2) = \frac{1}{4\pi}\int \dd^2x \sqrt{g}\,h_1^{\mu\nu}h_{2\mu\nu}\,,
	\\
	(\lambda_1,\lambda_2) & = \frac{1}{2\pi} \int \dd^2x \sqrt{g}\,\lambda_1^{\mu}\lambda_{2\mu}\,, \qquad [b,c] = \int \dd^2x \sqrt{g}\,b^{\mu}c_{\mu}\,,
\end{split}
\end{align}
The bosonic part of the field space metric, expressed in terms of the coefficients $\Phi_{\ell m}, \omega_{\ell m}$, etc., is diagonal in $\ell m$, i.e.
\beq
	\dd s^2_{\rm field} = \sum_{\ell=0}^{\infty} \sum_{m=-\ell}^{\ell} \dd s^2_{\ell ,m}\,.
\eeq
For $\ell=0$ and $\ell\geq 2$ we have
\begin{align}
\begin{split}
	\dd s^2_0 & = \frac{1}{2\pi} \left( \dd \Phi_{\ell=0}^2 + \dd \omega_{\ell=0}^2\right)\,,
	\\
	\dd s^2_{\ell\geq 2,m} & = \frac{1}{2\pi} \left( \dd\Phi_{\ell m}^2 + \dd\omega_{\ell m}^2 - 2 \sqrt{\ell(\ell+1)} \dd\omega_{\ell m}\dd V_{\ell m} + 2(\ell^2+\ell-1)\dd V_{\ell m}^2 \right.
	\\
	& \qquad \qquad \qquad \left.+ (\ell-1)(\ell+2)\dd\tilde{V}_{\ell m}^2 + \dd\lambda_{\ell m}^2 + \dd\tilde{\lambda}_{\ell m}^2\right)\,,
\end{split}
\end{align}
The $\ell=1$ subspace must be treated more carefully. Using that
\beq
	\sqrt{g_{\ell=0}} = \frac{1}{2\pi}\,, \quad \sqrt{g_{\ell\geq 2,m}} = \frac{(\ell-1)(\ell+2)}{(2\pi)^3}\,,
\eeq
and accounting for the ghosts the path integral measure becomes
\begin{align}
\begin{split}
	\frac{[\dd\Phi][dh]}{\text{diff}} &=\frac{1}{2\pi} \dd\Phi_{\ell=0}\dd\omega_{\ell=0} \times \left( \text{measure on }\ell=1\text{ modes}\right)
	\\
	&\times \prod_{\ell=2}^{\infty} \prod_{m=-\ell}^{\ell}\left(  \frac{(\ell -1)(\ell+2)}{(2\pi)^3} \dd\Phi_{\ell m}\dd\omega_{\ell m}\dd\lambda_{\ell m}\dd\tilde{\lambda}_{\ell m}\dd V_{\ell m} \dd\tilde{V}_{\ell m}\dd b_{\ell m}\dd c_{\ell m}\dd\tilde{b}_{\ell m}\dd\tilde{c}_{\ell m}\right)\,.
\end{split}
\end{align}
The bosonic part of the $\ell=1$ subspace has degenerate directions, which we can take to be $V_{\ell=1}$ and $\tilde{V}_{\ell=1}$. Moreover the gauge-fixing on this subspace must be treated carefully. As we will see shortly, the $\ell=1$ modes of $\omega$ are pure gauge and can be fixed using a Faddeev-Popov complex with $\lambda_{\ell=1}$ and $(b_{\ell=1},c_{\ell=1})$. There are no parity-odd $\ell=1$ fluctuations of the metric, and the parity-odd $\ell=1$ Faddeev-Popov complex should be understood as really a division by the $SO(3)$ isometries of the round sphere saddle. Let us call the carefully normalized isometry volume $\mathcal{V}$. Then the $\ell=1$ measure is
\beq
	\left( \frac{[\dd \Phi][\dd h]}{\text{diff}}\right)_{\ell=1} =\frac{1}{\mathcal{V}} \prod_{m=-1}^{m=1} \frac{1}{(2\pi)^{3/2}} \dd \Phi_{\ell=1,m} \dd \omega_{\ell=1,m} \dd \lambda_{\ell=1,m} \dd b_{\ell=1,m} \dd c_{\ell=1,m}\,.
\eeq

We rotate the dilaton by $\pm 90^{\circ}$ (which does not introduce a phase), so that for $\ell \geq 2$ the fields $(\omega_{\ell},\lambda_{\ell},\tilde{\lambda}_{\ell})$ act as Lagrange multipliers. For $\ell\geq 2$ there is perfect cancellation between the modes, and so the one-loop determinant becomes
\beq
	Z_{1\text{-loop}} = Z_{\ell=0}Z_{\ell=1}\,, \qquad Z_{\ell=0} = \frac{1}{2}\,,
\eeq
where $2$ is the eigenvalue of the bosonic $\ell=0$ sector.

Now let us turn to the $\ell=1$ determinant. To evaluate it we must account for the residual gauge symmetry generated by solutions of $(\nabla^2 + 1)\xi_{\mu} = 0$. These are conformal Killing vectors
\beq
	\xi^{ \mu} =\sum_{a=1}^3 \left( K^a  \nabla^{\mu} X^a + R^a\varepsilon^{\mu\nu}\nabla_{\nu}X^a\right)\,,
\eeq
with $X^a$ Cartesian coordinates on the sphere $X^a = (\sin\theta\cos\psi,\sin\theta\sin\psi,\cos\theta)$. The $X^a$ are linear combinations of the $\ell=1$ scalar harmonics, and the corresponding ghosts are the $\ell=1$ vector harmonics of $b^{\mu}$ and $c_{\mu}$. The second term in the diffeomorphism generates the $SO(3)$ isometry, whose volume we must carefully compute using the path integral measure, while the first generates a transformation of $\omega$ through 
\beq
	\delta_{\sigma} \omega = \nabla_{\mu}\xi^{\mu}  = - 2 K^a X^a\,.
\eeq
That is, the $\ell=1$ modes of $\omega$ are pure diffeomorphism. We gauge-fix the $\ell=1$ modes of $\omega$ using $(\lambda,b,c)$, and trade the integration over the tilde'd $\ell=1$ modes for a careful division by the isometry volume $\mathcal{V}$. To enact the former we simply fix $\omega_{\ell=1}=0$ using the $\ell=1$ modes of $\lambda$, and use the $\ell=1$ modes of $b$ and $c$ as the compensating ghosts. This corresponds to appending to our quadratic action the gauge-fixing and ghost terms
\begin{align}
\begin{split}
	S_{\rm residual}& = i \sum_{m=-1}^1 \left( \int \dd^2x \sqrt{g}\,\lambda^{\mu} e_{\mu}^{\ell=1,m}\right)\left( \int \dd^2y \sqrt{g} \,\omega \,Y_{\ell=1,m}^*\right)
	\\
	& \qquad + \sum_{m=-1}^1\left( \int \dd^2x \sqrt{g} \,b^{\mu}e_{\mu}^{\ell=1,m}\right)\left( \int \dd^2y \sqrt{g}(-\nabla_{\mu}c^{\mu})Y_{\ell=1,m}^*\right)\,,
	\\
	& = \sum_{m=-1}^1 \left(  i \lambda_{\ell=1,m}^* \omega_{\ell=1,m} + \sqrt{2}\,b_{\ell=1,m}^* c_{\ell=1,m}\right),
\end{split}
\end{align}
where $e_{\mu}^{\ell m}$ are the parity-even vector harmonics
\beq
	e_{\mu}^{\ell m} = \nabla_{\mu} \frac{Y_{\ell m}}{\sqrt{\ell(\ell+1)}}\,,
\eeq
orthonormalized with respect to the ultralocal measure. So the $\ell=1$ determinant is
\beq
	Z_{\ell=1} = 2^{3/2}\times \frac{1}{\mathcal{V}}\times \frac{1}{U^{3/2}}\,.
\eeq
The $2^{3/2}$ comes from the proper treatment of the untilde'd $\ell=1$ fields above; $\frac{1}{\mathcal{V}}$ from the division by isometries; and $U^{-3/2}$ from the integral over the $\ell=1$ modes of the dilaton.

To evaluate $\mathcal{V}$ we follow the approach of~\cite{Anninos:2020hfj}.  We denote by $\mathcal{V}\equiv\mathcal{V}_{\rm PI}$ the volume of the residual isometry group measured with the metric induced from the same ultralocal path integral measure used above. This isometry group is $SO(3)$.\footnote{In \textbf{v1} of this manuscript we incorrectly identified the isometry group as $O(3)$, which gave $\mathcal{V}_{\rm there} = 2\mathcal{V}_{\rm here}$. In computing the sphere amplitude, we divide only by orientation-preserving diffeomorphisms, so the stabilizer is $SO(3)$ rather than $O(3)$. We thank H.~Z.~(Vincent) Chen for discussions on this point.} If $\dd s^2_{\rm canon}$ is the canonical metric on $SO(3)$ and $\dd s^2_{\rm PI}$ is the path-integral metric on the same group orbit, then
\beq
\mathcal{V} = \text{vol}_{\rm canon}(SO(3)) \left( \frac{\dd s^2_{\rm PI}}{\dd s^2_{\rm canon}} \right)^{\frac{1}{2}\dim SO(3)} .
\eeq
We use the canonical metric $\dd s^2_{\rm canon} = \sum_{a=1}^3 (\omega^a)^2$ on $SO(3)$, with $\omega^a$ the Maurer-Cartan one-forms normalized so that the corresponding one-parameter subgroups have period $2\pi$.  With this normalization, $\text{vol}_{\rm canon}(SO(3)) = 8\pi^2$.\footnote{Thinking of $SO(3)$ as $SU(2)/\mathbb{Z}_2$, and $SU(2)$ as the three-sphere, in our normalization this $\mathbb{S}^3$ has radius $r=2$ so that great circles in $SO(3)$ have length $2\pi$. This leads to $\text{vol}(SO(3)) = \frac{1}{2}\text{vol}(\mathbb{S}^3) = \frac{1}{2}(2\pi^2 r^3) = 8\pi^2$.}  Given~\eqref{E:ultralocal}, the metric on the space of rescaled diffeomorphisms (acting on rescaled metric fluctuations as $\delta_{\xi} h_{\mu\nu} = \nabla_{\mu}\xi_{\nu} + \nabla_{\nu}\xi_{\mu}$) is\footnote{See Appendix G of~\cite{Anninos:2020hfj} for a discussion of this fact. The precise normalization may also be deduced by deriving the Faddeev-Popov procedure for a single normalized mode.}
\beq
\label{E:diffMeasure}
\dd s^2_{\rm PI} = \frac{1}{2\pi} \int \dd^2x\sqrt{g}\,\xi^\mu\xi_\mu \,.
\eeq

Now let $q^a$ be local coordinates along the three rotational isometries, chosen so that $\dd s^2_{\rm canon} = \sum_{a=1}^3(\dd q^a)^2$ near the identity.  In terms of the rescaled fluctuation variables used above, an infinitesimal displacement along the group orbit is represented by
\beq
\xi^\mu = \frac{\dd q^a}{\sqrt{16\pi G}}\,
	\varepsilon^{\mu\nu}\nabla_\nu X^a ,
\eeq
with summation over $a$ understood.  Equivalently, the unrescaled diffeomorphism vector field is $\sqrt{16\pi G}\,\xi^\mu$, so that the generator with $\dd q^3=1$ is $\partial_\psi$.  Using
\beq
\int \dd^2x\sqrt{g}\, \nabla_\mu X^a\nabla^\mu X^b = \frac{8\pi}{3}\,\delta^{ab},
\eeq
we find
\beq
\dd s^2_{\rm PI} = \frac{1}{12\pi G}\, \sum_{a=1}^3(\dd q^a)^2 = \frac{1}{12\pi G}\,\dd s^2_{\rm canon}
\eeq
and thus $\frac{\dd s^2_{\rm PI}}{\dd s^2_{\rm canon}} = \frac{1}{12\pi G}$, giving us
\beq
\mathcal{V} = 8\pi^2 \left( \frac{1}{12\pi G} \right)^{3/2}.
\eeq

Putting the pieces together, $Z_{1\text{-loop}} = Z_{\ell=0}Z_{\ell=1} = \frac{1}{2}\, 2^{3/2}\, \frac{(12\pi G)^{3/2}}{8\pi^2}\, \frac{1}{U^{3/2}} = \frac{2}{\sqrt{2\pi}} \left( \frac{3G}{U} \right)^{3/2}$. Then in the one-loop approximation,
\beq
\label{E:Zsphere}
	\boxed{ Z_{\rm sphere} = \frac{2e^{2S_0}}{\sqrt{2\pi}} \left( \frac{3G}{U}\right)^{3/2}}
\eeq
A different normalization of measure would give the same result up to an additive renormalization of $S_0$. Note that at one-loop the amplitude only depends on the potential $\mathcal{U}$ through its curvature $U$ at the saddle $\Phi=0$.

Note that if $U<0$, we must further rotate the $\ell=1$ modes of $\Phi$ back, leading to a phase $(\pm i)^3$. (The particular phase can be fixed by equipping $1/G$ with an $i\varepsilon$ prescription as in~\cite{Maldacena:2024spf}.) In the next Subsection we will see that $U<0$ corresponds to the no-boundary wavefunction being a wrong-sign Gaussian distribution.

By the usual arguments the sphere exponentiates in the sum over closed geometries. So, to leading order in the genus expansion we have
\beq
\label{E:Zclosed}
	Z_{\rm closed} = \exp\left( \frac{2e^{2S_0}}{\sqrt{2\pi}}\left( \frac{3G}{U}\right)^{3/2} + O(e^{0\times S_0})\right)\,.
\eeq

\subsubsection{No-boundary norm}
\label{subsubsec:no-boundary-norm}

At $U=0$ we recover ordinary dS JT gravity, for which we know the asymptotic states and their inner product. At late time we can always put the metric and dilaton into the form
\beq
\label{E:asymptotic}
	\dd s^2 = -\dd t^2 + (e^{2t} + O(1)) \dd\theta^2\,, \qquad \Phi = \frac{\phi(\theta)}{2\pi} e^t + O(1)\,,
\eeq
with $\theta\sim\theta+2\pi$. The boundary condition is labeled by the dilaton $\phi(\theta)$. States fall into three superselection sectors: (i) those where $\phi(\theta)$ has no roots (the rootless sector), (ii) those where $\phi(\theta)$ has roots~\cite{Alonso-Monsalve:2024oii,Held:2024rmg}, and (iii) certain scaling limits where $\phi(\theta)$ has roots~\cite{Alonso-Monsalve:2024oii}. The no-boundary wavefunction has support only on the rootless sector, and within it we can always use a residual diffeomorphism to `straighten' out the dilaton so that $\phi(\theta) = \varphi$, a constant~\cite{Cotler:2023eza}. The one-universe component of the no-boundary state is then described by the wavefunction
\beq
	\Psi_{\rm HH}(\varphi) = \langle\!\langle \varphi| \text{HH}\rangle\!\rangle\,.
\eeq
In the topological expansion this is computed by summing over geometries with one asymptotic boundary that cap off smoothly in the interior and asymptote to~\eqref{E:asymptotic}.  Its leading contribution is the disk, giving $\Psi_{\rm HH}\propto e^{S_0}$.

The inner product of the asymptotic $|\varphi\rangle\!\rangle$ states can be obtained by two complementary methods. One is a JT analogue of the `short time' path integral that computes inner products in quantum field theory~\cite{Cotler:2019dcj}. The other is to form an integration measure, meaning a completeness relation of boundary states, in terms of an integral over boundary data. In this context the boundary data is a boundary einbein $e(\theta)$ and a weight-one scalar $\phi(\theta)$, modulo diff$\times$Weyl. 

Let us briefly explain this second derivation, which is similar in spirit to that proposed in the Appendix of~\cite{Maldacena:2019cbz}.\footnote{There the authors obtain the measure $\dd L/L$ by integrating over a boundary einbein modulo diffeomorphisms.  In the rootless sector this is precisely a gauge-fixed form of the quotient over $(e,\phi)$ modulo diff$\times$Weyl since fixing $\phi = \pm 1$ gives one copy of $\dd L/L$ on each connected component.  In the gauge $e = 1$, the same quotient is instead written as the signed measure $\dd\varphi/|\varphi|$, with $L = 2\pi/|\varphi|$.} From the late-time limit we find the boundary data $(e(\theta),\phi(\theta))$ with $e>0$. There are large 2d diffeomorphisms that act as diff$\times$Weyl on this data, at the infinitesimal level
\beq
	\delta_{\chi} e = (\xi e)' + \sigma e\,, \qquad \delta_{\chi} \phi = \xi \phi' + \sigma \phi\,.
\eeq
On the sector where $\phi(\theta)$ is rootless\footnote{We evaluate the analogous measure where $\phi$ has roots in the Appendix.} $\phi(\theta)$ has definite sign for all $\theta$, and we can completely fix the Weyl symmetry by setting $\phi = \pm 1$ depending on the sign. The Faddeev-Popov determinant associated with this fixing is 1. Then we have the integration over $e(\theta)$ mod diff, which reduces to the standard $\dd L/L$ measure with $L$ the length of the circle in units where $\phi = \pm 1$. From this point of view the rootless sector has two parts, the one with $\phi > 0$ and the one with $\phi < 0$, each with the measure $\dd L/L$. 

A more useful parameterization is to use the Weyl symmetry to fix $e(\theta)=1$, so that the length of the circle in this gauge is $2\pi$.  The remaining data is then a signed weight-one scalar $\phi(\theta)$ modulo diffeomorphisms.  In the rootless sector it can be brought to a constant, $\phi(\theta) = \varphi$ for $\varphi \in \mathbb{R}\setminus\{0\}$. This parameterization is equivalent to the previous one.  Starting from the gauge $e = 1$ and $\phi = \varphi$, a Weyl transformation to the gauge $\phi = \pm 1$ gives $e = \frac{1}{|\varphi|}$ and  $L = \int_0^{2\pi}\dd\theta\,e = \frac{2\pi}{|\varphi|}$.  Thus, as a positive measure, $\left|\frac{\dd L}{L}\right| = \frac{\dd|\varphi|}{|\varphi|}$. Including both connected components of the rootless sector, $\varphi > 0$ and $\varphi < 0$, gives
\beq
\int_{\phi>0}\frac{\dd L}{L} + \int_{\phi<0}\frac{\dd L}{L} = \int_{\mathbb{R}}\frac{\dd\varphi}{|\varphi|}\,.
\eeq
The resulting completeness relation or projector onto single-universe rootless states is
\beq
\label{E:universeCompleteness}
\Pi_1 = \int_{-\infty}^{\infty} \frac{\dd\varphi}{|\varphi|} |\varphi\rangle\!\rangle \langle\!\langle\varphi|\,,
\eeq
or equivalently
\beq
\langle\!\langle \varphi|\varphi'\rangle\!\rangle =  |\varphi|\delta(\varphi-\varphi')\,.
\eeq

Said another way, the single-universe contribution to the norm of a state can be expressed as
\beq
	\int \frac{[\dd e][\dd\phi]}{\text{diff}\times\text{Weyl}} |\Psi[e,\phi]|^2 = \int_{-\infty}^{\infty} \frac{\dd\varphi}{|\varphi|} |\Psi(\varphi)|^2 + (\text{rooted sectors})\,,
\eeq
where the integral on the left-hand side is taken over configurations on a single circle and we are using that gravity produces wavefunctions $\Psi[e,\phi]$ that are invariant under diff$\times$Weyl.

Since the finite-time completeness relation and its detailed contour prescription will be developed separately in~\cite{CTJSWIP1}, in the present work we evaluate the norm using its controlled late-time form.

Our strategy in what follows is to first calculate the disk contribution to the no-boundary wavefunction, i.e.~the approximate single-universe wavefunction, 
\beq
	\Psi_{\rm HH}(\varphi) = \langle\!\langle \varphi|\text{HH}\rangle\!\rangle \approx \int \frac{[\dd g][\dd\Phi]}{\text{diff}}\,e^{iS}
\eeq
where the gravitational path integral is taken over metrics on a topological disk that cap off smoothly in the past and asymptote to~\eqref{E:asymptotic} in the future. With the wavefunction in hand we can calculate its contribution to the norm using~\eqref{E:universeCompleteness}.

Because of the quadratic term in the dilaton potential this approach will only be approximate, and at sufficiently large time the interaction will alter the asymptotics.

We first identify the saddle point of the gravitational path integral that computes the wavefunction using that classical dilaton gravity is `integrable' for any potential $V$~\cite{Maxfield:2020ale, Witten:2020wvy}. We work at small $\varphi$. The metric can be easily found using the dilaton as a `time coordinate' with
\beq
\label{E:diskSaddle}
	\dd s^2 = -\frac{\dd\Phi^2}{W(\Phi)} + W(\Phi)\left( \frac{2\pi}{\varphi}\,\dd\theta\right)^2 \,, \qquad W(\Phi) = \int_{\Phi_c}^{\Phi}\dd\Phi'\,V(\Phi')\,.
\eeq
$W$ is the `prepotential.' Consider the `late time' (but not too late) scaling limit $W \approx \Phi^2$. In that limit we change coordinates as $t = \ln \!\left( \frac{2\pi \Phi}{\varphi}\right)$ so that the geometry and dilaton are approximately
\beq
\label{E:lateTime}
	\dd s^2 \approx -\dd t^2 + e^{2t} \dd\theta^2\,, \qquad \Phi \approx \frac{\varphi}{2\pi}\,e^t
\eeq
so that~\eqref{E:diskSaddle} obeys the correct late-time boundary condition. The geometry caps off smoothly at $\Phi = \Phi_c$, which fixes 
\beq
	\varphi = \mp i \pi  V(\Phi_c)\,,
\eeq
so that there is a complex time contour. Demanding that imaginary time always decreases along the contour and expanding around the dS JT solution yields for small $\varphi$
\beq
	\Phi_c = \frac{i}{2\pi}\varphi - \frac{U}{16\pi^2} \varphi^2 + O(\varphi^3)\,.
\eeq
The tree-level approximation to the wavefunction is $e^{iS}$ with $S$ the on-shell action of this solution, including boundary terms consistent with our boundary condition. To $O(\varphi^2)$ we require a $\Phi^2$ boundary term, and the result is\footnote{We can think of the boundary terms as redefinitions of asymptotic states $\langle\!\langle \varphi|$ by a phase.}
\beq
\label{E:Srenormalized}
	S =-i S_0 +  \frac{1}{16\pi G} \int \dd^2x \sqrt{-g}(\Phi R - V) - \frac{1}{8\pi G}\int \dd\theta \sqrt{\gamma} \,\Phi K +   \frac{1}{8\pi G}\int \dd\theta \sqrt{\gamma} \left(\Phi - \frac{U}{12}\Phi^2\right)\,,
\eeq
with a boundary at time $\Phi =\pm \Lambda$ (depending on the sign of $\varphi$) with $\gamma$ the induced metric on the $\Phi =\pm \Lambda$ slice. The action is
\beq
 \label{E:onShell}
	S =-i S_0 +  \frac{\varphi}{16\pi G} + \frac{i U}{192\pi^2 G}\,\varphi^2  + O\!\left( \frac{\varphi^3}{G}\right)\,.
\eeq	

It is instructive to obtain the same result by working directly in perturbation theory in $\varphi$ around the no-boundary saddle point of dS JT gravity. For this approach,
\begin{align}
\begin{split}
	\dd s^2 & = -\dd t^2 + \cosh^2(t) \left( 1 + \varphi h(t)\right)\dd\theta^2 + O(\varphi^2)\,,
	\\
	\Phi & = \frac{\varphi}{2\pi}\sinh(t)\left( 1 +\varphi f(t)\right)+ O(\varphi^3)\,,
\end{split}
\end{align}
where at $\mathcal{U}=0$ the time $t$ traverses the complex time contour beginning at $\frac{i \pi}{2}$ and ending at $t \to \infty$. Solving the linearized equations and imposing regularity at $t = \frac{i \pi}{2}$ fixes
\begin{align}
\begin{split}
	h & = \frac{U}{6\pi}\left( i-\sinh(t)\right)\,,
	\\
	f & = f_0 -\frac{U}{24\pi}\left( \frac{2}{\sinh(t)} + \sinh(t)\right)\,.
\end{split}
\end{align}
Note that to this order the solution only depends on the curvature $U$ of the potential at $\Phi=0$. To fix the constant $f_0$ we go to late time. After a shift $t\to t+\ln 2 -\frac{iU\varphi }{12\pi }$ we have for $e^t\gg 1$
\begin{align}
\begin{split}
	\dd s^2 &\approx -\dd t^2 + \left( e^{2t} - \frac{U}{6\pi}e^{3t}\varphi \right)\dd \theta^2 \,,
	\\
	\Phi & \approx \frac{\varphi}{2\pi}e^t \left( 1+\varphi \left( f_0 - \frac{i U}{12\pi}\right)\right) -\frac{U}{48\pi^2}\varphi^2 e^{2t} \,,
\end{split}
\end{align}
so that to fix the late-time boundary condition we take $f_0 = \frac{i U}{12\pi}$ and the late-time scaling limit means $e^t\gg 1 \gg |U \varphi |e^t$. It is in this intermediate late-time regime that we use the asymptotic dS JT states and their inner product.  The limit should therefore not be understood as $t \to \infty$ at fixed nonzero $U\varphi$: the condition $e^t \gg 1$ places us in the asymptotic dS JT region, while $|U\varphi| e^t \ll 1$ ensures that the deformation has not yet appreciably modified the asymptotic data or their overlap.  Since the norm integral below is dominated by $|\varphi|\sim\sqrt{G/U}$, such a window exists parametrically when $UG\ll 1$, which is consistent with the semiclassical approximation $G\ll 1$ and $U$ fixed.

Now the complex time contour runs between $t = \frac{i \pi}{2}$ and $t = (\text{real and large}) - \frac{i U\varphi}{12\pi}$. The action~\eqref{E:Srenormalized} of this perturbed solution is precisely~\eqref{E:onShell}.

In pure dS JT gravity the tree-level approximation to the no-boundary distribution is a constant since $\Psi_{\rm HH,dS} \sim e^{iS} = e^{S_0+\frac{i\varphi}{16\pi G}}$ is a pure phase. However, accounting for the interactions we find a Gaussian distribution in $\varphi$,
\beq
\label{E:distribution}
	|\Psi_{\rm HH}|^2 \sim \exp\!\left(2S_0 - \frac{U}{96\pi^2 G}\,\varphi^2 + O(\varphi^3)\right)\,.
\eeq
This distribution is right-sign for $U>0$, the same sign for which the sphere amplitude was real and positive.

We stress that~\eqref{E:distribution} is the universal result near $\varphi=0$ for the tree-level distribution in dilaton gravity for a potential with a sphere saddle at $\Phi = 0$ and curvature $U$ there. 

To evaluate the norm we require the wavefunction to one loop.  In JT gravity the one-loop prefactor is proportional to $\varphi^{3/2}$.  In the deformed model and at small $\varphi$, where the leading perturbation to the background is $O(U\varphi)$, the prefactor is
\beq
	\varphi^{3/2}\left(1 + O(U\varphi) \right)\,,
\eeq
so at leading order in the small-$\varphi$ saddle it is controlled by the JT limit.  The one-loop wavefunction of dS JT gravity is known from the Schwarzian path integral~\cite{Maldacena:2019cbz,Cotler:2019nbi}. In the conventions of~\cite{Cotler:2024xzz}, the boundary variable is related to ours by $\Phi_{\rm there} =  \frac{\varphi_{\rm here}}{16\pi G}$. A rescaling of the Schwarzian path integral measure by a constant $\alpha$ rescales the one-loop prefactor by $\alpha^{-3/2}$. We evaluate the disk wavefunction with the same normalization of the GPI measure~\eqref{E:ultralocal} as in our computation of the sphere amplitude. It corresponds to $\alpha = \frac{1}{8G}$, which we now show. Along the way, we will also see that the phase in the dS JT measure obtained in~\cite{Cotler:2024xzz} emerges naturally from the GPI measure.

Let us work in coordinates so that the disk saddle is the Poincar\'e disk in all-minus signature,
\beq
\label{E:allMinus}
	\dd s^2 = - \frac{4}{(1-r^2)^2}(\dd r^2 + r^2 \dd\theta^2)\,,  \qquad \Phi_0 = i \frac{\varphi}{2\pi}\frac{1+r^2}{1-r^2}\,. 
\eeq
To see that this obeys the correct boundary condition~\eqref{E:lateTime} use the change of coordinates $r = \tanh\!\left( \frac{t+\ln 2- \frac{i \pi}{2}}{2}\right)$ with $t$ real Lorentzian time, and expand at large $t$. (The $-\frac{i \pi}{2}$ ensures that the complex time contour evolves down in imaginary time. The real late-time volume element $\sqrt{-g}$ becomes $i\sqrt{|g|}$ in terms of this metric.)

The action for quadratic fluctuations around the saddle is
\begin{align}
\begin{split}
	S & = S_{\rm tree} + S_2 + S_{\lambda} + S_{\rm FP} + O(\text{cubic})\,,
	\\
	S_{\rm tree} & =- i S_0 + \frac{\varphi}{16\pi G}\,,
	\\
	S_2 & = \frac{i}{16\pi G}\int \dd^2x \sqrt{|g|} \left( -\delta\Phi (\nabla^2 +2)\omega + \Phi_0\left( \frac{1}{4}h_{\mu\nu}h^{\mu\nu} - \frac{3}{8}h^2\right) \right)\,,
	\\
	S_{\lambda} & = \frac{i}{16\pi G} \int \dd^2x \sqrt{|g|}\,\lambda^{\mu}\left( \nabla^{\nu}h_{\mu\nu} - \frac{1}{2}\nabla_{\mu} h\right)\,,
	\\
	S_{\rm FP} & = \frac{i}{16\pi G}\int \dd^2x \sqrt{|g|}\,b^{\mu}(\nabla^2+1)c_{\mu}\,,
\end{split}
\end{align}
in terms of real fluctuations around the all-minus saddle,
\beq
	\Phi = \Phi_0+\delta\Phi\,, \qquad h_{\mu\nu} = g_{\mu\nu} \omega + \nabla_{\mu}V_{\nu}+\nabla_{\nu}V_{\mu}\,,
\eeq	
with $\nabla_{\mu}$ the covariant derivative constructed from~\eqref{E:allMinus}. We rotate the dilaton fluctuation, Lagrange multipliers, and ghosts by $\pm 90^{\circ}$ (all of which do not introduce a phase) so that the dilaton fluctuation and $\lambda^{\mu}$ act as Lagrange multipliers and the ghost determinant is manifestly positive-definite (as it ought). In order to ensure an apples-to-apples comparison between the sphere and one-boundary norm, we use the same normalization~\eqref{E:ultralocal} of the GPI we used in computing the sphere, now a real pairing on the Lorentzian-signature fluctuations. Around the all-minus saddle the pairing for the non-rescaled metric fluctuations is
\beq
	(h_1,h_2) = \frac{i}{4\pi} \frac{1}{16\pi G}\int \dd^2x \sqrt{|g|} \,h_1^{\mu\nu}h_{2\mu\nu}\,.
\eeq
Expanding the fluctuations into eigenmodes of the corresponding differential operators, the bosonic and ghost determinants cancel mode by mode apart from an exceptional sector of metric fluctuations, corresponding to the Schwarzian mode. Recall that there was a similar effect on the sphere where, in this normalization of measure, the modes with $\ell >1$ canceled.

The exceptional modes in question obey
\beq
	\nabla^{\mu}h_{\mu\nu} - \frac{1}{2}\nabla_{\nu} h = 0\,.
\eeq
They are the diffeomorphisms (the restriction to $|n|>1$ will be clear shortly)
\begin{align}
\begin{split}
	V^{\mu}\partial_{\mu} &= \frac{1}{4}\sum_{n \neq -1,0,1}\varepsilon_n e^{i n\theta} r^{|n|-2}\left( -i n r(1-r^2)(1+r^2+|n|(1-r^2)) \partial_r \right. 
	\\
	& \qquad \qquad\qquad\qquad \qquad \qquad \left. +(4r^2+|n|(1-r^4)+n^2(1-r^2)^2)\partial_{\theta}\, \right)\,,
\end{split}
\end{align}
where reality implies $\varepsilon_n^* = \varepsilon_{-n}$ and so $\text{Re}(\varepsilon_n) = \text{Re}(\varepsilon_{-n})$ and $\text{Im}(\varepsilon_n) = -\text{Im}(\varepsilon_{-n})$. These are large diffeomorphisms, taking the late time form
\beq
	V^{\mu}\partial_{\mu} \approx \sum_{n\neq -1,0,1} \varepsilon_n e^{in\theta}\partial_{\theta}\,.
\eeq
As vector fields they are non-normalizable, but the corresponding metric fluctuations are normalizable with
\beq
	(h,h)= \frac{i}{4\pi G}\sum_{n>1} n(n^2-1) |\dd\varepsilon_n|^2\,.
\eeq
The modes with $n=-1,0,1$ have vanishing measure -- these generate the $SO(2,1)$ isometries of the disk -- and so the GPI does not integrate over them, explaining the restriction to $|n|>1$. This leads to the measure
\beq
	\frac{[\dd h][\dd \Phi]}{\text{diff}} = \prod_{n>1} \left( -\frac{n(n^2-1)}{4\pi G}\dd\text{Re}(\varepsilon_n)\dd\text{Im}(\varepsilon_n)\right) \times (\text{modes that cancel})\,.
\eeq
This is the measure on the Schwarzian modes in~\cite{Cotler:2024xzz} with $\alpha = \frac{1}{8G}$ as claimed.

We can do better and reproduce the disk wavefunction here. The quadratic action evaluated on these exceptional fluctuations is
\beq
	S_2 =  \frac{\varphi}{8 \pi G} \sum_{n>1} n^2(n^2-1) |\varepsilon_n|^2\,,
\eeq
which comes entirely from the bulk part of the action. The one-loop disk wavefunction is (after picking appropriate steepest descent contours for the $\varepsilon_n$) then
\beq
	\Psi_{\rm HH,JT} \approx  e^{S_0 + i \frac{\varphi}{16\pi G}} \prod_{n>1}\left(\frac{2\pi i} {n \varphi}\right) \,,
\eeq
which under Zeta regularization becomes\footnote{This matches the dS JT result in~\cite{Cotler:2024xzz} $ \Psi_{\rm HH,JT} = \frac{e^{S_0}}{\sqrt{2\pi}}\, \alpha^{-3/2} \left( \frac{-i\varphi}{16\pi G} \right)^{3/2} \exp\!\left( \frac{i\varphi}{16\pi G} \right) $ with $\alpha = \frac{1}{8G}$.}
\beq
\Psi_{\text{HH,JT}}(\varphi) \approx \frac{e^{S_0}}{4\pi^2}(-i\varphi)^{3/2} \exp\!\left( \frac{i\varphi}{16\pi G}\right)\,.
\eeq

Putting the pieces together we arrive at the approximate one-loop wavefunction of dilaton gravity at small $\varphi$,
\beq
	\Psi_{\rm HH}(\varphi) \approx \frac{e^{S_0}}{4\pi^2}(-i\varphi)^{3/2} \exp\!\left( \frac{i\varphi}{16\pi G} - \frac{U}{192\pi^2 G}\,\varphi^2 \right)\,.
\eeq
The one-universe norm is then 
\beq
	\langle \!\langle \text{HH}|\Pi_1 |\text{HH}\rangle\!\rangle = \int_{-\infty}^{\infty} \frac{d\varphi}{|\varphi|} |\Psi_{\rm HH}(\varphi)|^2 \approx \frac{e^{2S_0}}{16\pi^4} \int_{-\infty}^{\infty}d\varphi\,\varphi^2 \exp\!\left( -\frac{U}{96\pi^2 G}\,\varphi^2\right)\,,
\eeq
giving the one-loop result
\beq
	\boxed{ \langle \!\langle \text{HH}|\Pi_1 |\text{HH}\rangle\!\rangle_{1\text{-loop}} =4\times  \frac{2e^{2S_0}}{\sqrt{2\pi}}\left( \frac{3G}{U}\right)^{3/2}}
\eeq
exactly four times the sphere amplitude~\eqref{E:Zsphere}.

Note that since $\varphi$ is Gaussian distributed with a width $\sqrt{G/U}$, corrections to the prefactor of the wavefunction or non-Gaussianities in the exponent will generate corrections to the norm that are suppressed by positive powers of $G$ relative to the boxed result. 

We also observe that the distribution for $\varphi$ is right-sign for $U>0$. For $U<0$ the distribution is wrong-sign, and the distribution for boundary data is maximized somewhere else. If we were to try to account for the unstable extremum at $\varphi=0$ through a saddle-point integration, we would rotate $\varphi$ by $\pm 90^{\circ}$ and find an imaginary contribution to the norm going as $(\pm i)^{3}$.

So far we have considered the one-universe contribution to the norm. There are also multi-universe contributions. On $n$-universe states $|\varphi_1,\hdots \varphi_n\rangle\!\rangle$ in JT gravity we have the overlap (with $n$-universe states orthogonal to $m$-universe states when $n\neq m$)~\cite{Cotler:2019dcj}
\beq
	\langle \!\langle \varphi_1,\hdots \varphi_n|\varphi_1',\hdots \varphi_n'\rangle\!\rangle = \left|\prod_{i=1}^n\varphi_i\right| \sum_{\sigma \in S_n}\delta(\varphi_1 - \varphi_{\sigma(1)}') \hdots \delta(\varphi_n-\varphi_{\sigma(n)}')\,,
\eeq
where from the point of view of the short-time path integral the $n!$ terms come from $n!$ combinations of short cylinders connecting initial and final boundary conditions. So multi-universe states behave like identical bosons. The $n$-universe completeness relation (or projector) on rootless states then reads
\beq
	\Pi_n =\frac{1}{n!} \int\left(  \prod_{i=1}^n\frac{\dd\varphi_i}{|\varphi_i|}\right)|\varphi_1,\hdots,\varphi_n\rangle\!\rangle\langle\!\langle \varphi_1,\hdots,\varphi_n|\,.
\eeq
This result can also be inferred from the rootless part of $(e,\varphi)$ on $n$ disjoint circles, with the $1/n!$ arising as a symmetry factor from division by a $S_n$ permutation of the $n$ circles.

The $n$-universe contribution to the norm is then
\beq
	\langle \!\langle \text{HH}|\Pi_n|\text{HH}\rangle\!\rangle = \frac{1}{n!}\int \left( \prod_{i=1}^n \frac{\dd\varphi_i}{|\varphi_i|}\right)|\Psi_{\rm HH}(\varphi_1,\hdots,\varphi_n)|^2\,,
\eeq
where to leading order in the genus expansion we have
\beq
	\Psi_{\rm HH}(\varphi_1,\hdots,\varphi_n) \approx \prod_{i=1}^n \Psi_{\rm HH}(\varphi_i)\,,
\eeq
and so
\beq
	\langle\!\langle \text{HH}|\Pi_n|\text{HH}\rangle\!\rangle \approx \frac{\langle \!\langle \text{HH}|\Pi_1|\text{HH}\rangle\!\rangle^n}{n!}\,.
\eeq
Summing up the contribution from $n$-universe states means that the one-universe contribution exponentiates
\beq
	\langle \!\langle \text{HH}| \text{HH}\rangle\!\rangle =\sum_{n=0}^{\infty} \langle\!\langle \text{HH}|\Pi_n|\text{HH}\rangle\!\rangle \approx  \exp\left( \langle \!\langle \text{HH}|\Pi_1|\text{HH}\rangle\!\rangle\right)\,,
\eeq
giving
\beq
	\langle \!\langle \text{HH}| \text{HH}\rangle\!\rangle \approx \exp\!\left(4\times  \frac{2e^{2S_0}}{\sqrt{2\pi}}\left( \frac{3G}{U}\right)^{3/2}\right)\,.
\eeq
For reference this equals $\langle\!\langle \text{HH}|\text{HH}\rangle\!\rangle \approx \exp\left( 4 Z_{\rm sphere}\right)$, whereas the sum over closed geometries~\eqref{E:Zclosed} is $Z_{\rm closed} \approx \exp\left( Z_{\rm sphere}\right)$.

\subsection{Comparison with Einstein gravity}
\label{subsec:Einstein}

In Einstein gravity in $d$ dimensions the sphere amplitude is, schematically,
\beq
	Z_{\text{sphere},\,1\text{-loop}} =e^{\mathcal{S}}\ \times  i^{d+2}\times \mathcal{D}\times  \frac{\mathcal{S}^{-d(d+1)/4}}{\text{vol}(SO(d+1))}\,,
\eeq
with $\mathcal{S} = \frac{\text{vol}(\mathbb{S}^{d-2})}{4G}$ the tree-level horizon entropy of the static patch and $\mathcal{D}$ a positive one-loop constant that has been evaluated for a particular choice of renormalization in~\cite{Anninos:2020hfj}. The factor of $e^{\mathcal{S}}$ is the tree-level approximation to the result; the factor of $i^{d+2}$ is Polchinski's phase, coming from the steepest descent contour of integration over the conformal mode; $\mathcal{D}$ comes from the integration over the nonzero modes in the problem; and the last ratio comes from the treatment of the residual $SO(d+1)$ isometry of the saddle. In contrast, the one-loop approximation to the late-time norm, a GPI whose saddle point is a round sphere $\mathbb{S}^{d-1}$, computed in our previous work~\cite{Cotler:2025gui} is
\beq
	\langle\!\langle \text{HH}|\Pi_1|\text{HH}\rangle\!\rangle_{1\text{-loop}}  = e^{\mathcal{S}} \times 1 \times \widetilde{\mathcal{D}} \times \frac{\mathcal{S}^{-d(d+1)/4}}{\text{vol}(SO(d,1))}\,,
\eeq
where $\widetilde{\mathcal{D}}$ is a positive one-loop constant. The factor of $e^{\mathcal{S}}$ is again the tree-level approximation; there is no phase; the constant $\widetilde{\mathcal{D}}$ comes from the integration over nonzero modes; and the last ratio comes from the treatment of a residual $SO(d,1)$ conformal isometry of the late-time norm.

So, in Einstein gravity, we have
\beq
	Z_{\rm sphere} \neq \langle\!\langle \text{HH}|\Pi_1|\text{HH}\rangle\!\rangle\,,
\eeq
as in deformed JT gravity, although here the mismatch is more dramatic. Indeed, the right-hand side vanishes to one-loop, and in fact to any order in the gravitational perturbation theory~\cite{Cotler:2026wlk}.

Before going on, we note that in $d>3$ the late-time norm is likely nonzero non-perturbatively, namely with the leading corrections being non-perturbatively suppressed relative to the tree-level answer $e^{\mathcal{S}}$. The argument is as follows. The late-time norm can be expressed as an integral over conformal classes of metrics on the late-time sphere $\mathbb{S}^{d-1}$. In $d=3$ the integration domain is a single point, with vanishing measure, while in $d>3$ there are regions in the domain where the sphere has been ``squashed'' a finite amount. In these regions the wavefunction at tree-level has
\beq
	|\Psi[\gamma_{\rm squash}]|^2 \sim  \exp\!\left(\mathcal{S} - \frac{\Delta I[\gamma_{\rm squash}]}{G}\right)\,,
\eeq
for a finite cost $\Delta I$ at fixed squashing, i.e.~the distribution on late-time data is non-perturbatively suppressed relative to $\exp(\mathcal{S})$. In that region, however, the configurations have no conformal isometry and therefore carry an $O(1)$ measure.

The late-time norm is an integral over two complexified disks, conjugates of each other, glued together at late time. Topologically the total spacetime is a $d$-sphere. For this reason we infer that the Lorentzian overlap is computing a different sphere amplitude, call it $\widetilde{\mathcal{Z}}_{\rm sphere}$, which unlike the Euclidean one, is non-negative as befitting a norm, but which vanishes in perturbation theory thanks to a noncompact stabilizer $SO(d,1)$. So, the Euclidean sphere amplitude is not the norm defined by the late-time measure.

What if we compute the norm at earlier times? In future work~\cite{CTJSWIP1} we construct a version of the late-time overlap that applies at finite bulk time. Consider taking the time all the way to $t=0$, where we evaluate the wavefunction on the equator of the Euclidean hemisphere that prepares the state. Then $\widetilde{\mathcal{Z}}_{\rm sphere}$ becomes a pairing of hemisphere wavefunctions. Under the assumption that the Euclidean amplitude can be cut along the equator, we conclude that the Lorentzian and Euclidean amplitudes are different pairings of the same wavefunction with itself. We have some preliminary and direct evidence that this is true.

That is, the difference between the Euclidean and Lorentzian amplitudes indicates two different pairings computed by the GPI.

We also expect two different pairings in deformed dS JT gravity. We suspect the mismatch is ultimately due to different redundancies in the two problems. In the sphere we divide by the $SO(3)$ isometry, whereas in the norm there is an analogue of the division by $SO(d,1)$, corresponding to the $SO(2,1)$ residual isometry of the configuration $e=1$, $\varphi=0$, the maximum of the tree-level probability distribution $|\Psi(\varphi)|^2$. However the no-boundary wavefunction has joint zero modes of the einbein and dilaton, which can be fixed by the $SO(2,1)$ residual stabilizer to give the finite measure $\dd\varphi/|\varphi|$. 

\subsection{Comments on inflation and dimensional reduction}
\label{subsec:inflationDimensionalReduction}

In some ways dS JT gravity is analogous to Einstein gravity coupled to an inflaton. In the two-dimensional theory the late-time boundary data include a dilaton profile $\phi(\theta)$, and the no-boundary norm is obtained by integrating over this profile, together with the boundary einbein, modulo diff$\times$Weyl.  In this limited sense the dilaton profile plays a role reminiscent of a scalar field profile at future infinity.  Quadratic deformations of the dilaton potential produce a probability distribution for the zero mode $\varphi$, much as a scalar potential produces a probability distribution for matter data in higher-dimensional cosmology.

This analogy is imperfect. In deformed dS JT gravity the dilaton is part of the gravitational constraint system, and so is not a propagating matter field on top of a fixed de Sitter background.  Its boundary value $\phi(\theta)$ is a weight-one scalar under the Weyl transformations inherited from large two-dimensional diffeomorphisms, which is why the rootless quotient can be reduced either by fixing $\phi = \pm 1$ or, equivalently, to the signed measure $\dd\varphi/|\varphi|$.

In slow-roll backgrounds, however, the inflaton shifts under Weyl rescalings. Nevertheless, slow-roll backgrounds share one important feature with deformed JT: there are joint zero modes in the inflaton and metric that can be used to fix the noncompact part of the residual $SO(d,1)$ isometry in the late-time measure, as will be discussed in~\cite{CIMWIP1}. The end result is a finite late-time measure, with no division by a noncompact stabilizer. This measure includes an integral over the homogeneous $\ell = 0$ mode of the inflaton, $\chi_0$, which plays a similar role to $\varphi$ in deformed dS JT.

For deformed dS JT, the quadratic term in the potential produces a Gaussian distribution around the sphere saddle, and so at small $\varphi$ the late-time distribution is $\frac{\dd\varphi}{|\varphi|}\,|\Psi_{\rm HH}^{\rm JT}(\varphi)|^2 \sim\dd\varphi\,\varphi^2\exp\!\left( -\frac{U}{96\pi^2G}\varphi^2\right)$.  As we emphasized above, the quadratic deformation is important since at $U = 0$ the tree-level dS JT wavefunction is only a phase, whereas for $U > 0$ the no-boundary distribution is normalizable near $\varphi = 0$.

By contrast, in the usual slow-roll approximation the inflaton potential is expanded over a finite rolling range as
\beq
	V(\chi) = V_* + V'_*(\chi - \chi_*) + \cdots\,, \quad V'_* \neq 0\,.
\eeq
After fixing the boost zero modes, the distribution for $\chi_0$ is then governed schematically by the de Sitter entropy as a function of the local potential $|\Psi(\chi_0)|^2\sim \exp(\mathcal{S}_* - \kappa V'_*(\chi_0 - \chi_*) + \cdots)$, or  in the linearized slow-roll regime
\beq
	|\Psi(\chi_0)|^2 \sim \exp(-\kappa V'_*\chi_0)
\eeq
up to an irrelevant shift of $\chi_0$.  We note that this is exponential rather than Gaussian.  If the linear potential is extended indefinitely, the integral over $\chi_0$ diverges. There is a similar divergence in the Euclidean sphere amplitude with a linear potential. The late-time norm is therefore sensitive to how the slow-roll region is completed, in particular to the lower edge of the potential and to the exit from inflation.

There is a more direct connection between deformed dS JT gravity and Einstein gravity in the context of the $\mathbb{S}^1\times \mathbb{S}^{d-2}$ no-boundary wavefunction. Consider the $\mathbb{S}^2\times\mathbb{S}^2$ amplitude in $d=4$ Einstein gravity, together with the $\mathbb{S}^1 \times \mathbb{S}^2$ wavefunction. Working in units where the 4d action is  
\beq
	S = \frac{1}{16\pi G_4}\int \dd^4x \sqrt{-G}\,(R(G) - 6)\,,
\eeq
so that de Sitter solutions have unit radius, and reducing on a two-sphere with the ansatz
\beq
	\dd s_4^2 = \frac{1}{\sqrt{\Phi}}\dd s_2^2 + \Phi\,\dd\bar\Omega^2\,,
\eeq
gives the dilaton gravity theory
\beq
	S = \frac{1}{16\pi G_2}\int \dd^2x\sqrt{-g}\,(\Phi R(g) - V(\Phi))\,, \qquad \frac{1}{G_2} = \frac{4\pi}{G_4}\,, \qquad V(\Phi) = 6\sqrt{\Phi} - \frac{2}{\sqrt{\Phi}}\,.
\eeq
The Euclidean version has a sphere saddle at
\beq
	\Phi_0 = \frac{1}{3}\,, \qquad V(\Phi_0) = 0\,,
\eeq
which uplifts to the $\mathbb{S}^2\times\mathbb{S}^2$ saddle of the $d=4$ theory.  The same saddle continues to a dS$_2\times\mathbb{S}^2$ solution in real time.

Expanding the potential near $\Phi_0$, one finds
\beq
	V(\Phi) = \frac{2}{L_2^2}(\Phi - \Phi_0) - \frac{1}{2}U(\Phi - \Phi_0)^2 + O((\Phi - \Phi_0)^3)\,, \qquad L_2^2 = \frac{1}{3^{3/2}}\,, \qquad U = 18\sqrt{3}\,.
\eeq
Thus the curvature $U = -V''(\Phi_0)$ of the effective interaction is positive.  In the language of the previous deformed dS JT calculation, the fluctuation of the dilaton away from the saddle is therefore governed by a right-sign Gaussian.  The $dS_2\times\mathbb{S}^2$ solution itself has constant dilaton, corresponding to $\varphi = 0$ for the fluctuation variable.

This is consistent with the results of~\cite{Turiaci:2025xwi}.  With the identification $L_{\rm there} \propto 1/\varphi$, the large-$L$ regime corresponds to small $\varphi$.  The wavefunction found there behaves as $\varphi^3$, rather than as the $\varphi^{3/2}$ of pure two-dimensional dS JT.  The additional power is naturally attributed to the three isometries of the compactified $\mathbb{S}^2$.  The corresponding one-boundary norm then scales schematically as
\beq
	e^{-S_E}\int \dd\varphi\,|\varphi|^5\exp\!\left(-\frac{\#}{G}\varphi^2\right) \sim G^3 e^{-S_E}\,,
\eeq
where $S_E$ is the Euclidean action of the $\mathbb{S}^2\times\mathbb{S}^2$ saddle.  This matches the schematic scaling of the $\mathbb{S}^2\times\mathbb{S}^2$ amplitude, where the factor $G^3$ arises from quotienting by the six-dimensional compact isometry group $SO(3)\times SO(3)$, although our results for deformed JT above suggest the two differ by a multiplicative constant.

Although the dilaton $\Phi$ is non-negative in this dimensional reduction, since it is the warp factor for the transverse sphere, the integration variable $\varphi$ in the two-dimensional near-saddle analysis is a fluctuation around $\Phi_0 = 1/3$.  It can therefore have either sign.  This is consistent with the deformed dS JT computation above, where we integrated over both signs of $\varphi$.

\section{Beyond semiclassical overlaps}
\label{sec:obstructions}

The previous Section leaves us with two logically distinct questions. First, does the semiclassical Lorentzian overlap admit a non-perturbative extension that includes a sum over topologies and reduces to the ultralocal late-time pairing in the appropriate regime? Second, is there some, potentially different, sewing operation on no-boundary wavefunctions that reconstructs the Euclidean sum over closed geometries? The factor-of-four mismatch found already at sphere topology shows that these two questions cannot be identified: a completion that reduces to the semiclassical Lorentzian pairing necessarily inherits
\begin{align}
\langle\!\langle \text{HH}|\Pi_1|\text{HH}\rangle\!\rangle = 4Z_{\rm sphere}
\end{align}
at leading order in the genus expansion and at one loop, and therefore cannot simultaneously reproduce the Euclidean closed-geometry sum.

In Einstein gravity the first question is presently beyond us, but optimistically it is answerable in deformations of dS JT gravity. The second question can already be constrained rather strongly. The results below suggest that a sewing prescription which reconstructs the Euclidean gravitational path integral cannot, once topology is included, be a universal pairing determined solely by the intrinsic data on the cut.

The goal of this Section is to record several interrelated observations relevant to these two questions.

The first observation concerns a candidate topological completion of the Lorentzian overlap. To state it, we first express the semiclassical overlap used in the last Section as a sum over geometries. Consider the semiclassical overlap of $n$-universe states. The overlap comes from summing over ``short''\footnote{We will have more to say about the difference between ``short'' and ``long'' cylinders later in this Section.} cylinders that connect the universes prepared by the ket to those prepared by the bra. Any one contribution is a product of $n$ such cylinders, and there are $n!$ contributions corresponding to the $n!$ ways of linking universes in the ket to those in the bra.\footnote{In the measure, the inverse of the inner product, this becomes a factor of $1/n!$. It arises from division by large diffeomorphisms of the cut that permute the $n$ disjoint universes.} Physically this is a Bose symmetrization factor, so that multiple universes semiclassically behave like identical bosons. This is the overlap whose topological completion we would like to understand. In light of the result of the previous Section, it should not be identified with the sewing operation implicit in the Euclidean path integral.

\begin{figure}[t!]
\includegraphics[width=\textwidth]{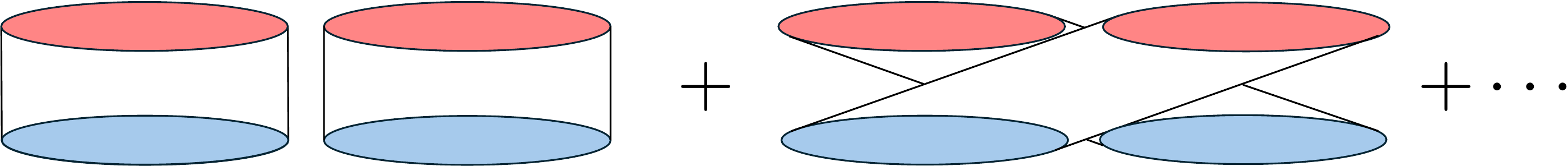}
\caption{
Depiction of a candidate topological completion of the Lorentzian overlap. Red circles denote bra universes and blue circles denote ket universes. At the semiclassical level, the two terms shown are the two ways of pairing two bra boundaries with two ket boundaries by ``short'' cylinders. More generally, the ellipsis denotes a sum over ``short'' geometries in which every connected component has at least one red boundary and at least one blue boundary.
}
\label{fig:red_blue}
\end{figure}

This result suggests a prescription extending the semiclassical overlap to one which sums over topologies. Namely, sum over ``short'' geometries that connect at least one universe in the ket to at least one universe in the bra. Implicitly this candidate requires treating the universes in the bra and those in the ket on a different footing. We can imagine assigning them colors, say blue to the ket universes and red to the bra universes, and require every connected geometry in the sum to end on at least one red boundary and one blue boundary.  See Fig.~\ref{fig:red_blue} for a depiction. The connectivity condition is well-defined topologically, but we do not know what ``short'' means beyond the cylinder. However even at this primitive stage it is clear that this proposal precludes contributions where the external states ``cap off'' or where bra universes are connected to each other and nothing else. 

The second observation is the most detailed analysis in this Section and appears in Subsection~\ref{subsec:MCGLargeDiff}. It concerns the distinct question of whether one can define a sewing pairing which reconstructs the Euclidean gravitational path integral. We find tension between the following two requirements:
\begin{enumerate}
\item The sewing pairing is state-independent and determined solely by the intrinsic data on the cut.
\item When applied to the no-boundary wavefunction, the pairing reproduces the sum over closed geometries, $\langle\!\langle \text{HH}|\text{HH}\rangle\!\rangle_{\rm sew} = Z_{\rm closed}$.
\end{enumerate}
The semiclassical Lorentzian pairing itself does not satisfy the second requirement: the factor-of-four discrepancy already demonstrates this at sphere topology. We therefore ask a different question here: could some other universal pairing of the same boundary wavefunctions reproduce the Euclidean path integral? We will see that in two-dimensional gravity there is a further obstruction: reconstructing the closed path integral requires global information that is not contained in the intrinsic data on the cut.

The third observation, discussed in Subsection~\ref{subsec:MM}, concerns the relation between the semiclassical overlap, its possible non-perturbative completion, and Marolf–Maxfield-like (MM) constructions. MM constructions have played an important role in understanding AdS quantum gravity, ensemble interpretations of gravitational path integrals, statistical averages over the black-hole microstate spectrum, and, more recently, closed universes. The topological sum suggested by the semiclassical overlap differs from an unrestricted MM-like sum already at low orders in the genus expansion. We also describe several features specific to Lorentzian de Sitter gravity, including in 3d, that any de Sitter version of such a construction must account for.

\subsection{State independence and Euclidean sewing}
\label{subsec:MCGLargeDiff}

At the level of pictures (and concretely in certain low-dimensional models), the late-time no-boundary wavefunction can be defined in the topological expansion using the GPI, namely integrating over smooth geometries that fill in late-time data. Similarly, the sum over closed manifolds can formally be defined with the GPI, again in the topological expansion. The result of the previous Section tells us that the actual semiclassical Lorentzian overlap is not the sewing operation that reconstructs the Euclidean GPI: already for two disks producing a sphere, the two differ by a factor of four.

Nevertheless, it is natural to ask whether the Euclidean path integral can be represented as some other pairing of the no-boundary wavefunctions, and what properties such a pairing would have. Our point here is that, assuming such a pairing exists, it cannot be determined solely by the data on the late-time surface. There is a tension between a state-independent sewing operation intrinsic to the cut and the requirement that sewing the no-boundary state reproduce $Z_{\rm closed}$.

\begin{figure}[t!]
\centering
\includegraphics[width=0.92\textwidth]{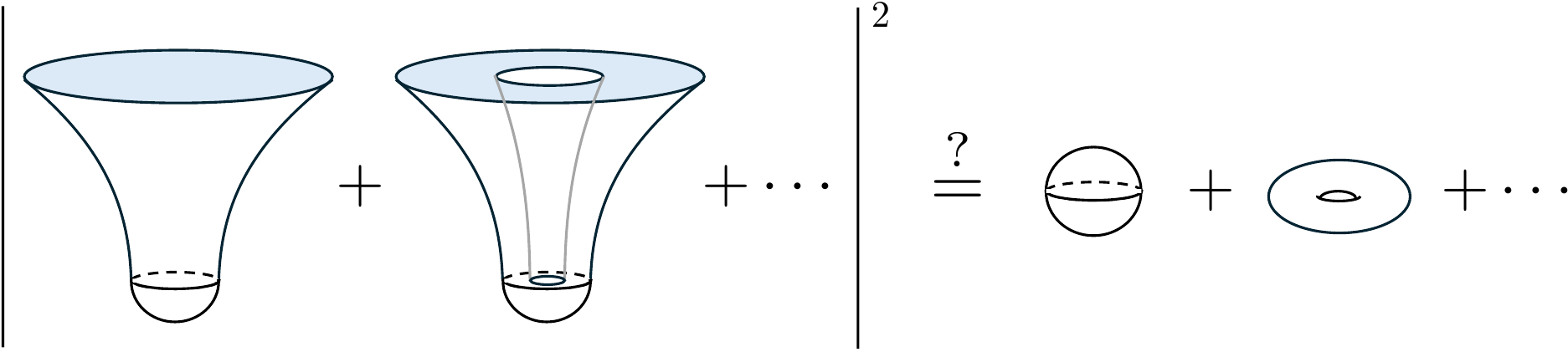}
\caption{
Gluing components of the no-boundary state produces a closed geometry together with a chosen decomposition into bra and ket pieces. Different cuts or decompositions can produce the same closed surface. A state-independent overlap acts on the boundary wavefunctions associated with a chosen decomposition, whereas the closed gravitational path integral assigns an amplitude to the closed surface without retaining this additional cutting data.
}
\label{fig:normsum}
\end{figure}

The basic problem is illustrated in Fig.~\ref{fig:normsum}. Cutting a closed surface produces more than the two states to be paired: it also introduces a separating curve and a decomposition into bra and ket regions. A fixed Hilbert-space pairing acts on the resulting boundary data, but it has no intrinsic knowledge of which closed surface or which alternative cut presentations those states came from. For the no-boundary norm to reproduce $Z_{\rm closed}$, the sewing operation must somehow forget this additional data. The examples below show that the semiclassical pairing does not accomplish this and, more strongly, that the sewing kernel required to reproduce closed amplitudes depends on the topology of the geometries away from the cut.

\subsubsection{Sewing depends on topology away from the cut}

Consider a closed manifold $M$ with a separating cut $\Sigma$ so that $M=M_+\cup_{\Sigma}M_-$, where $M_-$ and $M_+$ each have boundary $\Sigma$.\footnote{Our discussion comes with a caveat that is often left implicit in statements about ``gravitational path integrals.'' We assume that the path integral over metrics on the closed manifold $M$, and those on the pieces obtained after cutting along $\Sigma$, are meaningfully defined, so that one can speak of their respective contributions to $Z_{\rm closed}$ and to the no-boundary wavefunction. There are few regimes in which this is presently under quantitative control: chiefly saddle-point expansions, as in the previous Section, and certain low-dimensional models (perhaps including 3d gravity). The mapping-class-group effects discussed below are exact properties of off-shell amplitudes in two-dimensional gravity. We regard them as suggestive, rather than established, features of the much less understood off-shell amplitudes of higher-dimensional gravity.} With the appropriate orientation and conjugation assignments, the gravitational path integrals on $M_-$ and $M_+$ contribute respectively to the no-boundary ket and bra. The resulting wavefunctions depend on the gravitational data induced on $\Sigma$.

To reconstruct the GPI on $M$, one must sew these wavefunctions together by integrating over the data on $\Sigma$. But this is not enough. Cutting $M$ introduced additional information: a distinguished separating surface and a decomposition of $M$ into two pieces. The GPI on $M$ does not retain this marking. Sewing must therefore forget the marking that comes with introducing a cut. In reassembling $M$ we must divide by large diffeomorphisms that move the cut, and their action depends on the topology of $M$ away from $\Sigma$. This already suggests that the required sewing operation cannot be determined by the intrinsic data on $\Sigma$ alone.  A closely related issue with gravitational sewing in two-dimensional gravity was emphasized in Refs.~\cite{MooreSeibergStaudacher:1991,OkuyamaSakai:2021}, where it was observed that sewing at fixed moduli does not in general commute with integration over moduli space.

More explicitly, let
\beq
\Gamma(M)=\pi_0{\rm Diff}(M)\,,\qquad
\Gamma(M,\Sigma)={\rm Stab}_{\Gamma(M)}([\Sigma])\,,
\eeq
where $\Gamma(M,\Sigma)$ preserves the isotopy class of the cut. A fixed-cut construction naturally divides only by the mapping classes that preserve this marking, i.e.~by $\Gamma(M,\Sigma)$, whereas the GPI on $M$ divides by the full mapping class group $\Gamma(M)$. The orbit
\begin{equation}
\faktor{\Gamma(M)}{\Gamma(M,\Sigma)}
\end{equation}
describes cut presentations related by large diffeomorphisms of the closed manifold.

Passing from the fixed-cut amplitude to the closed amplitude must therefore account for this orbit. An integration measure intrinsic to $\Sigma$, however, has no knowledge of which other cuts are equivalent to it inside $M$. The mapping-class-group quotient thus identifies cut presentations using global information that a pairing determined solely by the cut data cannot contain.\footnote{In the semiclassical examples of the last Section, both the full and marked mapping class groups are trivial, so this feature is irrelevant there.}

It is important not to conflate this mapping class group obstruction with the factor-of-four mismatch found in the previous Section. For the sphere cut into two disks, the relevant mapping class groups are trivial. The factor of four is instead a finite one-loop mismatch associated with the different residual gauge quotients entering the Euclidean sphere amplitude and the Lorentzian late-time pairing. The mapping-class-group effect discussed here is an additional obstruction that appears once nontrivial topology is included. Thus even if one were to repair the sphere normalization by hand, the higher-topology sewing problem would remain.

We can make this statement precise in dilaton gravity. For example, Mirzakhani obtained~\cite{Mirzakhani:2006WP} the exact gluing formula in the context of hyperbolic surfaces where the cut $\Sigma$ is geodesic. The analogue of the no-boundary wavefunction is the Weil-Petersson volume of moduli space for a surface $M_-$ with boundary $\Sigma$, which depends on the geodesic length $L$ of the boundary, whose Weil-Petersson measure is $L\dd L$. In this context the gluing formula involves an integral over $L$ with measure $L\dd L \,K(L)$ where the kernel $K$ depends on the surfaces being sewn together. This kernel implements the division by large diffeomorphisms that move the cut.

This mismatch between the marked and unmarked amplitudes can be computed in dS JT gravity~\cite{Cotler:2024xzz}. Let $V_{g,n}(L_1,\ldots,L_n)$ denote the Weil-Petersson volume polynomial, where we omit the overall topological factor $e^{S_0\chi}$. In the $p$-basis (which is the Fourier transform of the `position basis' labeled by the renormalized dilaton $\Phi$; see~\cite{Cotler:2024xzz} for details) the local one-universe inner product, the analogue of the semiclassical overlap, is
\beq
\label{E:oneuniverse}
\langle \Psi|\Phi\rangle_{p} = \int_0^\infty \dd p\,\Psi^*(p)\Phi(p)\,.
\eeq
This measure is the local Hilbert-space measure.  It is not, by itself, the quotient measure on a closed moduli space.

In this basis the disk component of the de Sitter no-boundary state is distributional,
\beq
\Psi^{\rm dS}_{0,1}(p) = \delta'(p - 1)\,.
\eeq
The higher-genus one-boundary contributions take the form
\beq
\Psi^{\rm dS}_{g,1}(p) = (-1)^{3g - 2}\,V_{g,1}(2\pi i\sqrt p)\Theta(p)\,.
\eeq
Now let us glue the genus-$g$ one-boundary component to the disk using the inner product in~\eqref{E:oneuniverse}.  For $g \geq 2$,
\beq
\label{E:localgluing1}
\langle \!\langle \text{HH}_{g,1}|\text{HH}_{0,1}\rangle\!\rangle_{p} = (-1)^{3g - 2}\int_0^\infty \dd p\,V_{g,1}(2\pi i\sqrt p)\delta'(p - 1) = -(-1)^{3g - 2}\,\partial_p V_{g,1}(2\pi i\sqrt p)\big|_{p = 1}\,.
\eeq
To simplify this, we can use the Do-Norbury removable-cone relation $V_{g,0} = \frac{1}{2\pi i(2g - 2)}\,\partial_L V_{g,1}(L)\big|_{L = 2\pi i}$~\cite{Do:2006WP}. Since $L = 2\pi i\sqrt p$, one has $\partial_p L|_{p = 1} = i\pi$, and therefore $\partial_p V_{g,1}(2\pi i\sqrt p)\big|_{p = 1} = i \pi \,\partial_L V_{g,1}(L)\big|_{L = 2\pi i} = -(2\pi)^2(g - 1)V_{g,0}$. Then we can simplify~\eqref{E:localgluing1} as
\beq
\langle \!\langle\text{HH}_{g,1}|\text{HH}_{0,1}\rangle\!\rangle_{p} = (2\pi)^2(g - 1)V_{g,0}\,,
\eeq
where we have absorbed the sign $(-1)^{3g - 2}$ into the definition of the genus-$g$ component. The local pairing therefore gives $(2\pi)^2(g - 1)V_{g,0}$ rather than $V_{g,0}$.  The mismatch is a finite factor that depends on the genus of the surface being capped. Since this genus is information about the filling away from the cut, and is not part of the intrinsic boundary data labeled by $p$, no universal modification of the measure in~\eqref{E:oneuniverse} can remove the mismatch for every $g$. The sewing operation that reconstructs the closed amplitude depends on the topology of the state being sewn.

The genus dependence of this mismatch also shows why the factor-of-four mismatch at sphere topology cannot be resolved by a simple renormalization of the inner product by $1/4$. Even if one were willing to make such an ad hoc modification so that the disk--disk contraction reproduced the Euclidean sphere amplitude, the correction required in the higher-genus example above is proportional to $\frac{1}{(2\pi)^2(g-1)}$, and therefore depends on the topology of the state being capped. No topology-independent rescaling of the local one-universe measure can turn the Lorentzian inner product into the Euclidean sewing prescription.

There is a more severe mismatch when two non-disk components are contracted, in which case the orbit of the separating cut under the full mapping class group is infinite. With our conventions, the genus-$1$ one-boundary wavefunction is
\beq
\Psi^{\rm dS}_{1,1}(p) = \frac{\pi^2}{12}(p - 1)\Theta(p)\,,
\eeq
where we have absorbed a conventional overall sign into the definition of the wavefunction. The local pairing of the wavefunction with itself is
\beq
\langle \!\langle \text{HH}_{1,1}|\text{HH}_{1,1}\rangle\!\rangle_{p} = \frac{\pi^4}{144}\int_0^\infty \dd p\,(p - 1)^2 = \infty\,.
\eeq
Restoring the topological weights, the divergence occurs at order $e^{-2S_0}$.  The corresponding closed genus-two Weil-Petersson volume is finite,
\beq
V_{2,0} = \frac{43\pi^6}{2160}\,.
\eeq
Topologically, gluing two single-holed tori along their boundaries gives a genus-two surface.  The local contraction, however, first chooses a separating curve and then integrates over the asymptotic momentum $p$ using the one-universe Hilbert-space measure.  The large diffeomorphisms preserving that curve form only a subgroup of the full genus-two mapping class group.  By contrast, $V_{2,0}$ is the Weil-Petersson volume of the closed genus-two moduli space, which is finite because it divides by the full mapping class group.

\subsubsection{Multiplicity of decompositions}
\label{subsec:CombinatoricsGluing}

The previous Subsection considered a fixed decomposition $M=M_+\cup_\Sigma M_-$ and the mismatch between the mapping class group of $M$ and the subgroup preserving the cut. There is a related but distinct redundancy, stemming from the fact that a single closed manifold can often be assembled from components in many different ways.\footnote{We thank M.Van Raamsdonk for discussions on this point.} This gives a further obstruction to any state-independent sewing prescription intended to reproduce $Z_{\rm closed}$: gluing no-boundary wavefunctions together naturally retains a choice of decomposition into bra and ket components, whereas the closed gravitational path integral does not.

This is a combinatorial problem that we isolate in a two-dimensional topological toy model that retains only the Euler-characteristic weight of each connected surface. Let $\psi_n =\bar{\psi}_n= e^{S_0(2-n)}$ denote the connected ket and bra amplitudes of a genus-zero surface with $n$ boundaries, with $S_0\gg 1$. Thus $n=1$ is a cap, $n=2$ is a cylinder, $n=3$ is a pair of pants, and so on. A sphere contributes $e^{2S_0}$ to $Z_{\rm closed}$, so the leading term in the logarithm of the closed partition function is
\beq
	\ln Z_{\rm closed} =e^{2S_0} + O(e^{0\times S_0})\,.
\eeq

This overcounting is already visible for a single sphere. One way to build a sphere comes from gluing a ket cap to a bra cap. But the same sphere can also be obtained by gluing a genus-zero ket surface with $n$ boundaries to $n$ bra caps, or by more complicated assemblies involving several connected components on both sides. So there are infinitely many contributions in the norm that produce the same sphere.

Thus gluing no-boundary wavefunctions produces a sum over closed surfaces together with a choice of decomposition into bra and ket components. To reproduce the closed-manifold path integral, the overlap must forget this additional choice. We now make this statement explicit in the toy model.

Let $|{\rm HH}_0\rangle\!\rangle$ denote the genus-zero truncation of the full no-boundary state. Letting $|0\rangle$ be a fiducial, normalized 0-universe state, and letting $a$ and $a^{\dagger}$ (with $[a,a^{\dagger}]=1$) denote boundary annihilation and creation operators, it has the schematic form
\beq
|{\rm HH}_0\rangle\!\rangle = \exp\!\left(\sum_{n \geq 1}\frac{\psi_n}{n!}(a^\dagger)^n\right)|0\rangle\,.
\eeq
The factor $1/n!$ divides by permutations of the $n$ indistinguishable boundaries, while the exponential supplies the symmetry factors for disconnected products of connected components. This state includes many contributions; the term with $k_n$ copies of a sphere with $n$ disks removed is
\beq
\frac{1}{k_n!}\left(\frac{\psi_n}{n!}(a^\dagger)^n\right)^{k_n}|0\rangle\,,
\eeq
which carries a symmetrization factor $1/k_n!$ for the $k_n$ identical connected components and a factor $1/n!$ for the boundaries of each component.

The natural Fock-space norm pairs contributions with the same number of boundaries, gluing each ket boundary to a bra boundary. It is
\beq
\mathcal N_0 = \langle\!\langle{\rm HH}_0|{\rm HH}_0\rangle\!\rangle = \langle 0|\exp\!\left(\sum_{n \geq 1}\frac{\bar\psi_n}{n!}a^n\right)\exp\!\left(\sum_{n \geq 1}\frac{\psi_n}{n!}(a^\dagger)^n\right)|0\rangle\,.
\eeq
This norm accounts for every sphere assembled from genus-zero components, as well as some higher-genus closed surfaces. For example, gluing two cylinders along both pairs of boundaries produces a torus. To compute all higher-genus contributions, however, one must also include higher-genus terms in the no-boundary state.

We would like to compute the number of spheres produced by this overlap. In general, expanding the two exponentials gives bipartite gluing graphs.  Ket components are vertices on one side, bra components are vertices on the other side, and each edge glues one ket boundary to one bra boundary.  The connected part of the norm, $\log \mathcal N_0$, is the sum over connected bipartite gluing graphs.

For a connected graph built from genus-zero components, the characteristic of the resulting closed surface is
\beq
\chi = \sum_v(2 - \deg v) = 2V - 2E\,,
\eeq
where $V$ is the number of genus-zero components and $E$ is the number of boundary gluings.  The glued surface is a sphere when $\chi = 2$, which is equivalent to
\beq
E = V - 1\,.
\eeq
So spheres come from tree graphs. 

For a tree $T$, the contribution to $\ln \mathcal{N}_0$ is
\beq
W(T) = \frac{1}{|\operatorname{Aut}T|}\prod_{v\in V_K(T)}\psi_{\deg v}\prod_{w\in V_B(T)}\bar\psi_{\deg w}\,.
\eeq
The factor $|\text{Aut}\,T\,|^{-1}$ is what remains after combining the Fock-space symmetry factors with the contraction counting.  To see this, temporarily label the $r$ ket components and the $s$ bra components.  For a fixed labeled tree with ket degrees $d_i$ and bra degrees $e_j$, the Fock expansion gives
\beq
\frac{1}{r!\,s!}\prod_{i = 1}^r\frac{\psi_{d_i}}{d_i!}\prod_{j = 1}^s\frac{\bar\psi_{e_j}}{e_j!}\,.
\eeq
The number of contractions realizing this labeled tree is
\beq
\prod_{i = 1}^r d_i!\prod_{j = 1}^s e_j!\,,
\eeq
where boundary factorials cancel.  The labeled-tree contribution is therefore
\beq
\frac{1}{r!\,s!}\prod_{i = 1}^r\psi_{d_i}\prod_{j = 1}^s\bar\psi_{e_j}\,.
\eeq
Then summing over labeled trees with the factor $1/(r!\,s!)$ is equivalent to summing over unlabeled trees with weight $1/|\operatorname{Aut}T|$.

Using $\psi_n=e^{S_0(2-n)}$ and $E=V-1$ we see that every tree decomposition carries the same weight,
\beq
\prod_{v\in T}e^{S_0(2 - \deg v)} = e^{S_0(2V - 2E)} = e^{2S_0}\,.
\eeq
Now let us fix the number of ket components to be $r$ and the number of bra components to be $s$.  The number of labeled bipartite trees with $r$ vertices on one side and $s$ vertices on the other is $r^{s - 1}s^{r - 1}$.  After including the component symmetry factors, the fixed-$(r,s)$ contribution to $\ln \mathcal{N}_0$ is
\beq
W_{\mathbb{S}^2}^{r,s} = e^{2S_0}\frac{r^{s - 1}s^{r - 1}}{r!\,s!}\,.
\eeq
So the total sphere contribution to the no-boundary norm is
\beq
\ln \langle\!\langle \text{HH}|\text{HH}\rangle\!\rangle= e^{2S_0}\sum_{r,s \geq 1}\frac{r^{s - 1}s^{r - 1}}{r!\,s!} + O(e^{0\times S_0})\,.
\eeq
This sum is badly divergent. Along the diagonal $r=s=n$, the summand is  $\frac{n^{2n-2}}{(n!)^2}\sim\frac{e^{2n}}{2\pi n^3}$  which grows exponentially with $n$. This is manifestly different from $\ln Z_{\rm closed}=e^{2S_0}+O(e^{0\times S_0}).$ The mismatch arises because the no-boundary norm counts infinitely many decompositions of the same sphere. A non-perturbative overlap reproducing $Z_{\rm closed}$ must somehow forget this decomposition data.

\subsection{Comments on other candidate overlaps}
\label{subsec:MM}

In this Subsection we discuss features of Lorentzian de Sitter quantum gravity that obstruct a direct application of the Marolf--Maxfield philosophy, in which overlaps are defined by summing over geometries that fill specified boundary conditions. The factor-of-four mismatch provides the simplest example of the underlying issue: even when the two constructions involve the same sphere topology and the same no-boundary cap wavefunctions, the Euclidean gravitational path integral and the Lorentzian inner product implement different pairings. In Lorentzian de Sitter gravity there are further examples in which amplitudes with the same underlying topology compute manifestly different quantum-mechanical objects.

The basic issue is that, in a Lorentzian path integral, bras and kets, and especially past and future boundaries, play different roles. Bras and kets are related by suitable conjugation and orientation reversal operations, but the more pressing matter for us is that path integrals in non-topological theories involve an effective $i\varepsilon$ prescription when connecting the far past to the far future. A na\"{i}ve treatment of amplitudes, rooted in the classical intuition that $H=0$, would suggest that this distinction is unimportant, but there are concrete quantum gravity computations that indicate this is not the case.

These computations boil down to the statement that the global de Sitter amplitude, the ``long cylinder'' amplitude of dS quantum gravity, does not equal the ``short cylinder'' amplitude, the overlap corresponding to the semiclassical measure considered in the last Section. We recapitulate two simple examples of this effect.

First, consider the global dS amplitude of dS JT gravity, i.e.~the ``long cylinder.'' In the same conventions of Section~\ref{subsubsec:no-boundary-norm}, with the future state labeled by $\varphi_F$ and the past by $\varphi_P$, it is~\cite{Cotler:2024xzz}
\beq
	Z_{\rm cylinder}(\varphi_F,\varphi_P) = \frac{i}{2\pi} \frac{\sqrt{\varphi_F\varphi_P}}{\varphi_F - \varphi_P + i\varepsilon}\,,
\eeq
where we emphasize that an $i\varepsilon$ prescription is required to define the amplitude. In this convention the overlap of $\varphi$-states corresponding to the measure $d\varphi/|\varphi|$ discussed in Section~\ref{subsubsec:no-boundary-norm} is~\cite{Cotler:2024xzz}
\beq
	\langle\!\langle \varphi|\varphi'\rangle\!\rangle_{\rm ultralocal} = |\varphi|\delta(\varphi-\varphi')\,.
\eeq
This overlap is the ``short'' cylinder. The two clearly differ.\footnote{The ``long'' cylinder can be interpreted as a projection onto states of positive momentum $p$ conjugate to $\varphi$~\cite{Cotler:2023eza, Cotler:2024xzz}.} 

From this point of view, the long cylinder is better regarded as a transition amplitude between past and future states than as the overlap of boundary conditions on a common cut.

Another and more constraining example arises in pure 3d gravity. Let us consider cylinder amplitudes connecting sphere boundaries. The ``long'' cylinder amplitude has a saddle point, global dS$_3$, while the ``short'' cylinder amplitude corresponds to the ultralocal measure on boundary data. In the sector with sphere boundary conditions, the reduced boundary Hilbert space is one-dimensional, since the sphere has a unique conformal structure. Nevertheless, amplitudes computed by the GPI depend anomalously on the choice of conformal representative. Under Weyl rescaling, they transform like two-dimensional CFT partition functions with an effective central charge.

In particular, the global amplitude depends on the metric on the future boundary in the same way as a 2d CFT partition function with an effective central charge
\beq
	c= \frac{3i}{2G}+13\,.
\eeq
The imaginary part is a tree-level result, and the real shift of $+13$ arises at one-loop~\cite{Cotler:2019nbi}. The amplitude also depends on the metric on the past boundary in the same way as a 2d CFT with $c$. If the past and future metrics are identified, the global amplitude then carries total effective central charge $2c$, which includes (twice) the tree-level imaginary part.

By contrast, the ``short'' amplitude transforms with $c$ on the bra side and $\bar c$ on the ket side, giving a total effective central charge $c+\bar c=26$, the familiar result from the bosonic string.

Again, this tells us that the long cylinder amplitude is better regarded as part of a transition amplitude connecting a past Hilbert space to a future one. In particular, the global amplitude teaches us that the map from the far past to the far future involves a complex conjugation, whereby past kets of fixed boundary metric are behaving like future bras, in the sense that both transform under Weyl rescalings with central charge $c$, rather than conjugate central charges.

An adaptation of the Marolf-Maxfield philosophy to de Sitter gravity must account for these features. Indeed, if we define an overlap by filling in boundary conditions in all possible ways, the distinction between far past and far future tells us that we need to specify additional information. We can sum over
\begin{enumerate}
\item Geometries filling ket data in the far past and bra data in the far future.
\item Geometries filling bra and ket data on a common cut in the far past.
\item Geometries filling bra and ket data on a common cut in the far future.
\end{enumerate}
The first construction is naturally a transition amplitude. The second and third are more plausible candidates for an overlap. It is tempting to speculate that, once defined carefully in a non-topological theory, they reorganize into sums over the ``short'' geometries introduced at the beginning of this Section.

\section{Discussion}
\label{sec:discussion}

Let us summarize the main results of our manuscript, and some of the questions raised by them. Our main result concerns the semiclassical inner product of states at late time in de Sitter quantum gravity. We used this inner product to compute the late-time norm of the no-boundary state in deformations of dS JT gravity, to leading order in the genus expansion and to one-loop order. With a common ultralocal normalization of the gravitational path integral measure, we find $\langle\!\langle \text{HH}|\Pi_1|\text{HH}\rangle\!\rangle = 4 Z_{\rm sphere}$. The factor of four survives the careful treatment of the late-time measure, the one-loop disk wavefunction, and the residual gauge quotients on the two sides. Thus the Lorentzian completeness relation does not implement the sewing operation implicit in the Euclidean sphere path integral. At leading order in the genus expansion, the $n$-universe sectors factorize and the $S_n$ quotient supplies the usual $1/n!$, so that the Lorentzian norm exponentiates the one-universe answer, $\langle\!\langle \text{HH}|\text{HH}\rangle\!\rangle
\approx \exp(4Z_{\rm sphere})$, whereas the corresponding sum over disconnected Euclidean closed geometries is $Z_{\rm closed} \approx \exp(Z_{\rm sphere})$.

The analogous computation in Einstein gravity with positive cosmological constant leads to the same qualitative conclusion in a more dramatic form. In~\cite{Cotler:2025gui} we found that the perturbative late-time norm of the round-sphere sector vanishes because of a residual noncompact $SO(d,1)$ conformal stabilizer, whereas the Euclidean sphere amplitude is nonzero and carries Polchinski's phase. Thus both deformed dS JT gravity and Einstein gravity distinguish the Lorentzian pairing from the pairing implicit in the Euclidean path integral. In deformed JT the distinction is a finite factor of four at one loop; in Einstein gravity it separates a perturbatively vanishing norm from a nonzero complex Euclidean amplitude. This interpretation can be tested more directly at finite time. In forthcoming work~\cite{CTJSWIP1}, we pull the late-time overlap back to the equator and formulate both quantities as pairings of cap wavefunctions there. This should make precise how the Lorentzian and Euclidean constructions pair the same boundary data differently.

The factor-of-four result also sharpens the questions raised by the inclusion of topology. There are now two distinct problems. The first is to find a natural non-perturbative completion of the Lorentzian overlap itself. Such a completion should reduce to the ultralocal semiclassical pairing, and therefore should not be expected to satisfy $\langle\!\langle\text{HH}|\text{HH}\rangle\!\rangle=Z_{\rm closed}$.  Indeed, that equality already fails in the leading sphere sector. The second problem is whether the Euclidean gravitational path integral can nevertheless be represented as some other sewing pairing of no-boundary wavefunctions.

The examples in Section~\ref{sec:obstructions} place strong constraints on the latter possibility. In two-dimensional gravity, a fixed-cut pairing divides only by the mapping classes that preserve the cut, whereas a closed gravitational path integral divides by the full mapping class group. Reconstructing the closed amplitude therefore requires a sewing operation whose action depends on the topology of the geometries away from the cut. A related problem is that the same closed surface can arise from many decompositions into bra and ket components. In our topological toy model, the usual symmetrization factors do not identify these decompositions; instead, a na\"{i}ve Fock-space contraction contains a divergent sum over the infinitely many decompositions of a sphere. These are additional obstructions, beyond the finite one-loop mismatch already present at sphere topology, to interpreting the Euclidean path integral as a universal state-independent inner product intrinsic to the cut.

Another question concerns the relation between the semiclassical overlap and Marolf–Maxfield-like constructions defined by sums over bulk fillings. Lorentzian de Sitter gravity distinguishes amplitudes that have the same underlying topology. In dS JT gravity, the global de Sitter cylinder, a ``long'' cylinder connecting the far past to the far future, is a transition amplitude defined with an $i\varepsilon$ prescription. It does not equal the ultralocal ``short'' cylinder amplitude, the semiclassical overlap used in Section~\ref{sec:test}. Pure dS$_3$ gravity gives a sharper version of the same feature: the long global cylinder and the short overlap carry different effective anomaly coefficients. Thus a prescription to ``sum over all geometries that fill in boundary conditions'' does not by itself specify an inner product. A de Sitter version of the Marolf–Maxfield philosophy must additionally distinguish past from future through the Lorentzian time contour that selects which geometries contribute to transition amplitudes and which to overlaps.

Taken together, these observations leave two distinct questions open. The first is whether the Lorentzian semiclassical overlap has a natural non-perturbative completion. Such a completion should reduce to the ultralocal late-time pairing and retain the Lorentzian distinction between ``short'' overlaps and `long'' transition amplitudes; there is no reason for it to reproduce the Euclidean sum over closed geometries. The second is whether the Euclidean gravitational path integral admits a useful interpretation as a pairing of no-boundary wavefunctions. If such a Euclidean sewing pairing exists beyond fixed topology, our examples indicate that it must incorporate global information about cut presentations and decompositions, and therefore cannot be a universal state-independent pairing determined solely by the intrinsic boundary data. More broadly, our results suggest that the Lorentzian completeness relation and Euclidean gravitational sewing should be regarded as distinct operations.  This appears to resonate with recent works on OTOCs in the de Sitter static patch, which suggest that certain correlators on the Euclidean sphere may not have a Lorentzian interpretation in perturbative quantum gravity~\cite{Milekhin:2026tbi, Cui:2026bcd, Chen:2026boh, Harlow:2026pwe}.

\subsection*{Acknowledgements}

We would like to thank D.~Anninos, ChatGPT, H.Z.~(Vincent) Chen, V.~Gorbenko, V.~Ivo, J.~Maldacena, X.~Shi, G.~Turiaci, M.~Van Raamsdonk, A.~Vilar Lopez, and Z.~Yang for insightful discussions, and especially ChatGPT for comments on the draft. JC is supported by the Simons Collaboration on Celestial Holography, as well as a Fellowship from the Alfred P. Sloan Foundation. KJ is supported in part by an NSERC Discovery Grant. We acknowledge fruitful discussions during the workshop ``Observers, wormholes, and complex saddles in cosmology'', organized at the Bernoulli Center for Fundamental Studies (EPFL, Lausanne) from 18--22 May 2026.

\newpage

\appendix

\section{Boundary measure for rooted sector}
\label{app:rootedMeasure}

In the main text we used the boundary measure in the rootless sector of dS JT and its deformation.  Here we record what happens for smooth profiles with roots.  Recall that the boundary data are $(e(\theta),\phi(\theta))$ with $e(\theta) > 0$ modulo diff$\times$Weyl, and the infinitesimal action of diff$\times$Weyl is $\delta e = (\xi e)' + \sigma e$ and $\delta\phi = \xi\phi' + \sigma\phi$. Fixing the Weyl symmetry by setting $e(\theta) = 1$, the residual transformations are diffeomorphisms accompanied by the compensating Weyl transformation $\sigma = -\xi'$.  They act on $\phi$ as
\beq
\label{E:rootedResidualAction}
	\delta_\xi\phi = \xi\phi' - \xi'\phi\,.
\eeq
Equivalently, after this gauge fixing $\phi(\theta)$ transforms as a vector field on the circle:
\beq
	\phi(\theta) \mapsto \phi_f(\theta) = \frac{\phi(f(\theta))}{f'(\theta)}\,.
\eeq
For a rootless profile this quotient has the signed invariant used in the main text.  As we explained, we can bring $\phi$ to a constant representative $\phi(\theta) = \varphi$, or equivalently fix $\phi = \pm 1$ and keep the circle length $L = 2\pi/|\varphi|$.  This gives the measure $\dd\varphi/|\varphi|$ after including both signs.

Now suppose that $\phi$ is smooth, is not identically zero, and has at least one zero.  Then the residual transformation generated by
\beq
	\xi = \phi\,, \quad \sigma = -\phi'
\eeq
leaves the gauge-fixed boundary data invariant.  Indeed,
\beq
	\delta_{\xi=\phi}\phi = \phi\phi' - \phi'\phi = 0\,,
	\qquad
	\delta e = \xi' + \sigma = 0\,.
\eeq
This is the infinitesimal statement that a vector field is invariant under its own flow.  At finite flow time $s$, if $f_s$ is generated by $\phi$,
\beq
	\frac{\dd}{\dd s}f_s(\theta) = \phi(f_s(\theta))\,,
\eeq
then we have
\beq
	\frac{\phi(f_s(\theta))}{\partial_\theta f_s(\theta)} = \phi(\theta)\,,
\eeq
and so $f_s$ is a stabilizer of the profile.

For rootless $\phi$, this stabilizer is compact since the flow goes around the circle and returns to the identity after the period
\beq
	T = \int_0^{2\pi}\frac{\dd\theta}{\phi(\theta)}
\eeq
up to orientation.  For rooted $\phi$, the zeroes are fixed points of the flow.  On each interval between neighboring zeroes, the flow moves monotonically between the endpoints, and no nonzero finite flow time gives the identity diffeomorphism.  Thus a smooth rooted profile has a noncompact stabilizer $H_\phi \supset \mathbb R$. The special profile $\phi = 0$ has an even larger stabilizer.

In the ordinary unmarked quotient this stabilizer must be divided out.  Therefore a smooth rooted representative carries the residual factor
\beq
	\frac{1}{\Vol(H_\phi)} \supset \frac{1}{\Vol(\mathbb R)}\,.
\eeq
Equivalently, the Faddeev-Popov operator for the residual diffeomorphism quotient has the zero mode $\xi = \phi$.  Unless the state or observable contains a compensating distributional factor supported on the rooted stratum, the unmarked rooted contribution to the ordinary boundary measure vanishes
\beq
	\int_{\rm smooth\ rooted}\frac{[\dd e][\dd\phi]}{\text{diff}\times\text{Weyl}}\,F[e,\phi] = 0
\eeq
for smooth gauge-invariant $F$.

That said, there can still be useful rooted-sector states.  One may obtain a nonzero rooted-sector pairing by adding extra data that fixes the noncompact flow, for example by marking a point on an interval between roots, or by considering distributional states whose normalization compensates the $\Vol(\mathbb R)$.  Such constructions are different from the unmarked one-universe completeness relation used in the no-boundary norm.  In the computation in the main text the Hartle-Hawking wavefunction has support only in the rootless sector, so these rooted subtleties do not enter.

\bibliography{refs}
\bibliographystyle{JHEP}

\end{document}